\documentclass[article,12pt]{article} 
\usepackage{graphics,epsfig,amsmath,amsfonts,amssymb,color,wrapfig,colortbl,amsthm}

\usepackage{algorithm}
\usepackage{algorithmic}

\usepackage{graphicx}
\usepackage{amsmath}
\usepackage{tabularx}

\usepackage{subfigure}
\usepackage[english]{babel} % English
\usepackage{graphicx,amssymb,amstext,amsmath}
\usepackage{float}
\usepackage{amsfonts}
\usepackage{bbm}
\usepackage{bm}
\usepackage{verbatim}
\usepackage{subfigure}
\usepackage{framed}
\usepackage{hyperref}

\usepackage{flushend}
\usepackage{multirow}
\usepackage{makecell}
\usepackage{tabu}
\usepackage{mathtools}
\usepackage{amsmath}
\usepackage{bbm}
\usepackage{bm}

\usepackage{colortbl}

\definecolor{MYCOLOR0}{rgb}{0.92,0.92,0.92}
\definecolor{MYCOLOR}{rgb}{1,1,0}
\definecolor{MYCOLOR2}{rgb}{0.5,1,0.5}

\definecolor{MYCOLOR3}{rgb}{0.88,1,1}

\title{The Recycling Gibbs Sampler for Efficient Learning} 

\author{L. Martino$^{\star\Diamond}$, V. Elvira$^\top$, G. Camps-Valls$^\star$ \\
{\small$^\star$   Image Processing Laboratory (IPL), Universitat de Valencia, (Spain). }\\
{\small $^\top$ T\'el\'ecom ParisTech, Universit\'e Paris-Saclay. (France),} \\
{\small$^\Diamond$  Universidad Carlos III de Madrid,  Legan\'es (Spain).}
}
\date{}
 
\begin{document}

\maketitle

\thispagestyle{empty}

\begin{abstract}
Monte Carlo methods are essential tools for Bayesian inference. Gibbs sampling is a well-known Markov chain Monte Carlo (MCMC) algorithm, extensively used in signal processing, machine learning, and statistics, employed to draw samples from complicated high-dimensional posterior distributions. The key point for the successful application of the Gibbs sampler is the ability to draw efficiently samples from the full-conditional probability density functions. Since in the general case this is not possible, in order to speed up the convergence of the chain,  it is required to generate auxiliary samples whose information is eventually disregarded.  In this work, we show that these auxiliary samples can be recycled within the Gibbs estimators, improving their efficiency with no extra cost. This novel scheme arises naturally after pointing out the relationship between the standard Gibbs sampler and the chain rule used for sampling purposes. Numerical simulations involving simple and real inference problems confirm the excellent performance of the proposed scheme in terms of accuracy and computational efficiency. In particular we give empirical evidence of performance in a toy example, inference of Gaussian processes hyperparameters, and learning dependence graphs through regression. 
\newline
\newline
%\cite{Djuric03,Gordon93,Martino15PF}. It has been widely accepted as a valid measure of effective sample size. 
\end{abstract} 

%%%%%%%%%%%%%%%%%%%%%%%%%%%%%%%%%%%%%%%%
\section{Introduction}
%\label{sec-intro}
\begin{flushright}
{\em `Reduce, Reuse, Recycle'}\\
{\em The Greenpeace motto}
\end{flushright}

Monte Carlo algorithms have become very popular over the last decades~\cite{Liu04b,Robert04}.
Many practical problems in statistical signal processing, machine learning and statistics, demand fast and accurate procedures for drawing samples from probability distributions that exhibit arbitrary, non-standard forms~\cite{Andrieu2003,Fitzgerald01,ReadLuca2014}, \cite[Chapter 11]{Bishop}. One of the most popular Monte Carlo methods are the families of Markov chain Monte Carlo (MCMC) algorithms~\cite{Andrieu2003,Robert04} and particle filters~\cite{Bugallo07,Djuric03}. The MCMC techniques generate a Markov chain (i.e., a sequence of correlated samples) with a pre-established target probability density function (pdf) as invariant density~\cite{Liu04b,Liang10}. 
%The two most widely applied MCMC approaches are the Metropolis-Hastings (MH) algorithm and the Gibbs sampler~\cite{Liu04b,Robert04}.

The Gibbs sampling technique is a well-known MCMC algorithm, extensively used in the literature in order to generate samples from multivariate target densities, drawing each component of the samples from the full-conditional densities~\cite{Chen16,Koch07,Kotecha99,Goudie16,Lucka16,Zhang16}.\footnote{In general these full-conditionals are univariate. Nevertheless, block-wise Gibbs sampling approaches where several random variables are updated simultaneously, have been proposed to speed up the convergence of the Gibbs sampler~\cite{roberts1997updating}. However, unless direct sampling from the multi-variate full-conditionals is feasible, these approaches still result in an increased difficulty of drawing samples and a higher computational cost per iteration. Furthermore, the performance of the overall algorithm can decrease if the blocks are not properly chosen, especially when direct sampling from the multivariate full-conditionals is unfeasible~\cite{Liu04b,Liang10,Robert04}. The novel recycling scheme can also be used in the block approach. }
%
%When the multivariate target can be easily factorized into univariate conditional pdfs, 
In order to draw samples from a multivariate target distribution, the key point for the successful application of the standard Gibbs sampler is the ability to draw efficiently from the univariate conditional pdfs~\cite{Liu04b,Robert04}.
The best scenario for Gibbs sampling occurs when specific direct samplers are available for each full-conditional, e.g. inversion method or, more generally, some transformation of a random variable~\cite{Devroye86,Robert04}. 
Otherwise, other Monte Carlo techniques, such as rejection sampling (RS) and different flavors of the Metropolis-Hastings (MH) algorithms, are typically used {\it within} the Gibbs sampler to draw from the complicated full-conditionals. 
The performance of the resulting Gibbs sampler depends on the employed {\it internal} technique, as pointed out for instance in~\cite{Cai08,Gilks95,MartinoA2RMS,FUSS}. 

In this context, some authors have suggested to use more steps of the MH method within the Gibbs sampler~\cite{Muller91,Gelfand93,Fox12}. Moreover, other different algorithms have been proposed as alternatives to the MH technique~\cite{Cai08,Koch07,Shao13}. For instance, several automatic and self-tuning samplers have been designed to be used primarily {\it within-Gibbs}: the adaptive rejection sampling (ARS)~\cite{Gilks92derfree,Gilks92}, the griddy Gibbs sampler~\cite{ritter1992griddyGibbs}, the FUSS sampler~\cite{FUSS}, the Adaptive Rejection Metropolis Sampling (ARMS) method~\cite{Gilks95,CorrGilks97,Meyer08,Zhang16}, and the Independent Doubly Adaptive Rejection Metropolis Sampling (IA$^2$RMS) technique~\cite{MartinoA2RMS}, just to name a few. 

Most of the previous solutions require performing several MCMC steps for each full-conditional in order to improve the performance, 
%for instance allowing the adaptation of the parameters of the method) 
although only one of them is considered to produce the resulting Markov chain because the rest of samples play the mere role of auxiliary variables. Strikingly, they require an increase in the computational cost that is not completely paid off: several samples are drawn from the full-conditionals, but only a subset of these generated samples is employed in the final estimators. 
In this work, we show that the rest of generated samples can be directly incorporated within the corresponding Gibbs estimator. We call this approach the {\it Recycling Gibbs (RG) sampler} since {\it all} the samples drawn from each full-conditional can be used also to provide a better estimation, instead of discarding them. 

The consistency of the proposed RG estimators is guaranteed, as will be noted after considering the connection between the Gibbs scheme and the chain rule for sampling purposes~\cite{Devroye86,Robert04}. In particular, we show that the standard Gibbs approach is equivalent (after the burn-in period) to the standard chain rule, whereas RG is equivalent to an alternative version of the chain rule presented in this work as well. RG fits particularly well combined with adaptive MCMC schemes %, employed in order to draw from the full-conditional pdfs, 
where different internal steps are performed also for adapting the proposal density, see e.g.~\cite{Gilks95,MartinoA2RMS,Meyer08,Zhang16}. The novel RG scheme allows us to obtain better performance without adding any extra computational cost. This will be shown through intensive numerical simulations. First, we test RG in a simple toy example with a bimodal bivariate target. We also include experiments for hyper-parameter estimation in Gaussian Processes (GPs) regression problems with the so-called {\it automatic relevance determination} (ARD) kernel function~\cite{Bishop}. Finally, we apply the novel scheme in real-life geoscience problems of dependence estimation among bio-geo-physical variables from satellite sensory data. The MATLAB code of the numerical examples is provided at \url{http://isp.uv.es/code/RG.zip}.

The remainder of the paper is organized as follows. Section~\ref{BaySect} fixes notation and recalls the problem statement of Bayesian inference. The standard Gibbs sampler and the chain rule for sampling purposes are summarized in Section~\ref{SGsect}, highlighting their connections. In the same section, we then introduce an alternative chain rule approach, which is useful for describing the novel scheme. The RG technique proposed here is formalized in Section~\ref{NovelSect}. Sections~\ref{SIMU} provides empirical evidence of the benefits of the proposed scheme, considering different multivariate posterior distributions. Finally, Section~\ref{ConclSect} concludes and outlines further work.

%%%%%%%%%%%%%%%%%%%%%%% %%%%%%%%%%%%%%%%%%%%%%%%%%%%%%%%%%%%%%%%%%%%%%
\section{Bayesian inference} \label{BaySect}
%%%%%%%%%%%%%%%%%%%%%%% %%%%%%%%%%%%%%%%%%%%%%%%%%%%%%%%%%%%%%%%%%%%%%

Machine learning, statistics, and signal processing often face the problem of inference through density sampling of potentially complex multivariate distributions. In particular, Bayesian inference is concerned about doing inference about a variable of interest exploiting the Bayes' theorem to update the probability estimates according to the available information. Specifically, in many applications, the goal is to infer a variable of interest, ${\bf x}=[x_1,\ldots,x_{D}]\in  \mathbb{R}^{D}$, given a set of observations or measurements, ${\bf y}\in \mathbb{R}^{P}$. In Bayesian inference all the statistical information is summarized by means of the posterior pdf, i.e.,
\begin{equation}
	\bar{\pi}({\bf x})= p({\bf x}| {\bf y})= \frac{\ell({\bf y}|{\bf x}) g({\bf x})}{Z({\bf y})},
\label{eq:posterior}
\end{equation}
where $\ell({\bf y}|{\bf x})$ is the likelihood function, $g({\bf x})$ is the prior pdf and $Z({\bf y})$ is the marginal likelihood (a.k.a., Bayesian evidence).
In general, $Z({\bf y})$ is unknown and difficult to estimate in general, so we assume to be able to evaluate the unnormalized target function,
\begin{equation}
\pi({\bf x})=\ell({\bf y}|{\bf x}) g({\bf x}).
\label{eq:target}
\end{equation}
The analytical study of the posterior density $\bar{\pi}({\bf x}) \propto \pi({\bf x})$ is often unfeasible and integrals involving $\bar{\pi}({\bf x})$ are typically intractable. For instance, one might be interested in the estimation of
\begin{equation}
\label{MainInt}
I=\int_{\mathbb{R}^D} f({\bf x})\bar{\pi}({\bf x}) d{\bf x},
\end{equation}
where $f({\bf x})$ is a squared integrable function (with respect to $\bar{\pi}$). In order to compute the integral $I$ numerical approximations are typically required. Our goal here is to approximate this integral by using Monte Carlo (MC) quadrature~\cite{Liu04b,Robert04}. Namely, considering $T$ independent samples from the posterior target pdf, i.e., ${\bf x}^{(1)},\ldots,{\bf x}^{(T)} \sim \bar{\pi}({\bf x})$, we can write 
\begin{equation}
 {\widehat I}_T=\frac{1}{T} \sum_{t=1}^T f({\bf x}^{(t)})
\overset{p}{\longrightarrow}   I.
%\xrightarrow[]{p}
\end{equation}
This means that for the weak law of large numbers, ${\widehat I}_T$ converges in probability to $I$: that is, for any positive number $\epsilon>0$, we have $\lim\nolimits_{T\rightarrow \infty}\mbox{Pr}( |{\widehat I}_T-I |> \epsilon)=0 $. In general, a direct method for drawing independent samples from $\bar{\pi}({\bf x})$ is not available, and alternative approaches, e.g., MCMC algorithms, are needed. An MCMC method generates an ergodic Markov chain with invariant density $ \bar{\pi}({\bf x})$ (a.k.a., stationary pdf). Even though, the generated samples $\{{\bf x}^{(1)},\ldots,{\bf x}^{(T)}\}$ are then correlated in this case, ${\widehat I}_T$ %=\frac{1}{T} \sum_{t=1}^T f({\bf x}^{(t)})$ 
is still a consistent estimator. 
  
Within the MCMC framework, we can consider a block approach working directly into the $D$-dimensional space, e.g., using a Metropolis-Hastings (MH) algorithm~\cite{Robert04}, or a component-wise approach~\cite{HaarioCW,Johnson13,Levine05} working iteratively in different uni-dimensional slices of the entire space, e.g., using a Gibbs sampler~\cite{Liu04b,Liang10}.\footnote{There also exist intermediate strategies where the same subset of variables are jointly updated, which is often called the Blocked Gibbs Sampler.}  In many applications, and for different reasons, the  component-wise approach is the preferred choice. For instance, this is the case when the full-conditional distributions are directly provided or when the probability of accepting a new state with a complete block approach becomes negligible as the dimension of the problem $D$ increases. In the following section, we outline the standard Gibbs approach, and remark its connection with the chain rule method. The main notation and acronyms of the work are summarized in Table~\ref{tab:notation}.

\begin{table}[!t]
\centering
%\small
%\footnotesize
\caption{Main notation and acronyms of the work. }
\vspace{0.1cm}
	\begin{tabular}{|c|l||c|l|}
    \hline
%\begin{enumerate}
%\item 
 \cellcolor{MYCOLOR0} $D$ & \multicolumn{3}{l|}{Dimension of the inference problem, ${\bf x}\in\mathbb{R}^D$.} \\ 
 \cellcolor{MYCOLOR0} $T$ & \multicolumn{3}{l|}{Total number of iterations of the Gibbs scheme.} \\
 \cellcolor{MYCOLOR0}$M$ & \multicolumn{3}{l|}{Total number of iterations of the MCMC method inside the Gibbs scheme.} \\
  \cellcolor{MYCOLOR0}$t_b$ & \multicolumn{3}{l|}{Length of the burn-in period.} \\
\hline
\hline
 \cellcolor{MYCOLOR0}${\bf x}$ &\multicolumn{3}{l|}{Variable of interest; parameters to be inferred, ${\bf x}=[x_1,\ldots,x_{D}]$.}\\
 \cellcolor{MYCOLOR0}${\bf y}$ &\multicolumn{3}{l|}{Collected data: observations or measurements.} \\
 \cellcolor{MYCOLOR0}${\bf x}_{\neg d}$ & \multicolumn{3}{l|}{ ${\bf x}$ without the $d$-th component, i.e., $[x_1,\ldots,x_{d-1},x_{d+1},\ldots,x_D]$.} \\
 \cellcolor{MYCOLOR0}$x_{a:b}$ & \multicolumn{3}{l|}{The vector $x_{a:b}=[x_a,x_{a+1},x_{a+2},\ldots, x_b]$  with $b>a>0$.}\\
  \hline
  \hline
  \cellcolor{MYCOLOR0} $\bar{\pi}({\bf x})$ & \multicolumn{3}{l|}{Normalized posterior pdf $\bar{\pi}({\bf x})= p({\bf x}| {\bf y})$.} \\
  \cellcolor{MYCOLOR0} $\pi({\bf x})$ & \multicolumn{3}{l|}{Posterior function proportional to the posterior pdf, $\bar{\pi}({\bf x})\propto \pi({\bf x})$.} \\
   \hline
  \hline
   \cellcolor{MYCOLOR0}$\bar{\pi}_d(x_d|{\bf x}_{\neg d})$ &$d$-th full-conditional pdf. &  \cellcolor{MYCOLOR0}$p_d(x_d)$ &$d$-th marginal pdf. \\
  \hline
\hline
 \cellcolor{MYCOLOR0}SG& Standard Gibbs.  &  \cellcolor{MYCOLOR0}TRG &  Trivial Recycling Gibbs.  \\
 \cellcolor{MYCOLOR0}MH & Metropolis-Hastings. & \cellcolor{MYCOLOR0} MRG & Multiple Recycling Gibbs. \\
\hline 
\end{tabular}
\label{tab:notation}
\end{table}
%\multicolumn{8}{|c|}{Sets}

%%%%%%%%%%%%%%%%%%%%%%%%%%%%%%%%
\section{Gibbs sampling and the chain rule method} \label{SGsect}
%%%%%%%%%%%%%%%%%%%%%%%%%%%%%%%%

This section reviews the fundamentals about the standard Gibbs sampler, reviews the recent literature on Gibbs sampling when complicated full-conditional pdfs are involved, and points out the connection between GS and the chain rule. A variant of the chain rule is also described, which is related to the novel scheme introduced in the next section.

%%%%%%%%%%%%%%%%%%%%%%%%%
\subsection{The Standard Gibbs (SG) sampler}
%%%%%%%%%%%%%%%%%%%%%%%%%
The Gibbs sampler is perhaps the most widely used algorithm for inference in statistics and machine learning \cite{Chen16,Koch07,Goudie16, Robert04}. Let us define ${\bf x}_{\neg d} := [x_1,\ldots,x_{d-1},x_{d+1},\ldots,x_D]$ and introduce the following equivalent notations 
\begin{eqnarray*}
{\bar \pi}_d(x_d|x_1,\ldots,x_{d-1},x_{d+1},\ldots,x_D) = {\bar \pi}_d(x_d|x_{1:d-1},x_{d+1:D}) = {\bar \pi}_d(x_d|{\bf x}_{\neg d}).
 \end{eqnarray*}
In order to denote  the unidimensional full-conditional pdf of the component $x_d\in \mathbb{R}$, $d\in\{1,\ldots,D\}$, given the rest of variables ${\bf x}_{\neg d}$, i.e.
 \begin{eqnarray}
{\bar \pi}_d(x_d|{\bf x}_{\neg d}) = \frac{\bar{\pi}({\bf x})}{\bar{\pi}_{\neg d}({\bf x}_{\neg d})} = \frac{\bar{\pi}({\bf x})}{\int_{\mathbb{R}} \bar{\pi}({\bf x}) dx_{d}}.
 \end{eqnarray}
The density $\bar{\pi}_{\neg d}({\bf x}_{\neg d})=\int_{\mathbb{R}} \bar{\pi}({\bf x}) dx_{d}$ is the joint pdf of all variables but $x_d$. The Gibbs algorithm generates a sequence of $T$ samples, and is formed by the steps in Algorithm \ref{alg:gibbs}.
 Note that the main assumption for the application of Gibbs sampling is being able to draw efficiently from these univariate full-conditional pdfs $\bar{\pi}_d$. However, in general, we are not able to draw directly from any arbitrary full-conditional pdf. Thus, other Monte Carlo techniques are needed for drawing from all the $\bar{\pi}_d$.
%%%%%
\begin{algorithm}[htbp]
\caption{The Standard Gibbs (SG) algorithm.\label{alg:gibbs}}
  \begin{algorithmic}[1]
	\STATE{Fix $T$, $D$}
	\FOR{$t=1,\ldots,T$}
		\FOR{$d=1,\ldots,D$}
			\STATE{Draw $x_d^{(t)}\sim\bar{\pi}_d(x_d|x_{1:d-1}^{(t)},x_{d+1:D}^{(t-1)})$.}		
		\ENDFOR
		\STATE{Set ${\bf x}^{(t)}=[x_1^{(t)},x_2^{(t)},\ldots,x_D^{(t)}]$.}
	\ENDFOR
\end{algorithmic}
\end{algorithm}

\iffalse
%\begin{algorithm}
\begin{framed}
%\begin{algorithmic}
\begin{itemize}
\item For $t=1,\ldots,T$: 
\begin{enumerate}
\item For $d=1,\ldots,D$:
%\State 
\begin{enumerate}
\item  Draw $x_d^{(t)}\sim\bar{\pi}_d(x_d|x_{1:d-1}^{(t)},x_{d+1:D}^{(t-1)})$. 
%\State
\end{enumerate}
\item  Set ${\bf x}^{(t)}=[x_1^{(t)},x_2^{(t)},\ldots,x_D^{(t)}]$.
\end{enumerate}
\end{itemize}
%\EndFor
%\EndFor
\end{framed}
%\end{algorithmic}
%\end{algorithm}
\fi

%%%%%%%%%%%%%%%%%%%%%%% %%%%%%%%%%%%%%%%%%%%%%% %%%%%%%%%%%%%%%%%%%%%%%
\subsection{Monte Carlo-within-Gibbs sampling schemes}
\begin{figure}[htbp]
\centering
\includegraphics[width=15.4cm]{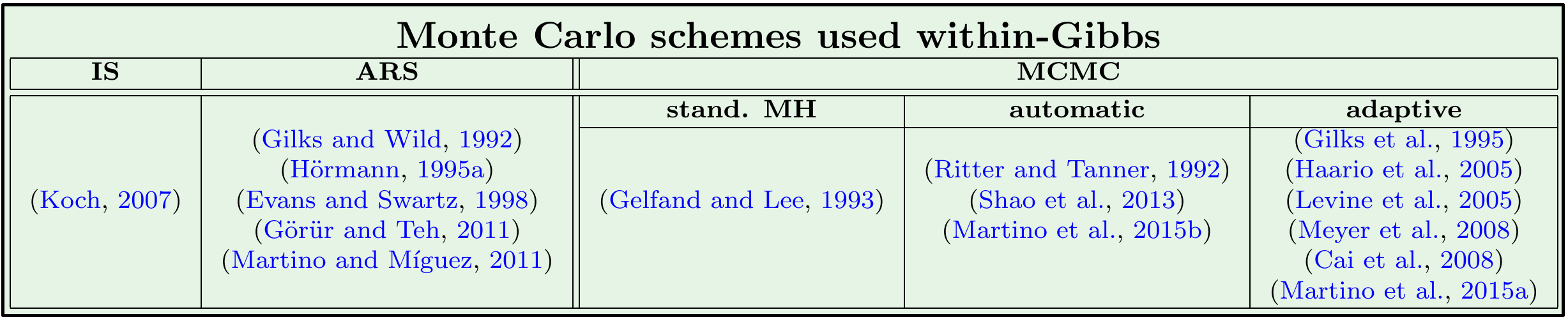} 
\caption{Summary of the main Monte Carlo algorithms which have been employed within the Gibbs sampling technique. }
\label{FigMonteCarloGibbs}
\end{figure}

%%%%%%%%%%%%%%%%%%%%%%%%%%%%%%%%%%%%%%%%%%%%%%%% %%%%%%%%%%%%%%%%%%%%
In many cases, drawing directly from the full-conditional pdf is not possible, hence the use of another Monte Carlo scheme is needed. Figure \ref{FigMonteCarloGibbs} summarizes the main techniques proposed in literature for this purpose.
In some specific situations, rejection samplers~\cite{Caffo02,Hormann02,Hormann07,Marrelec04,Tanizaki99} and their adaptive version, as the {\it adaptive rejection sampler} (ARS) \cite{Gilks92}, are employed to generate one sample from each $\bar{\pi}_d$ per iteration. Since the standard ARS can be applied only to log-concave densities, several extensions have been introduced \cite{Hoermann95,Gorur08rev,MartinoStatCo10}. Other variants or improvements of the standard ARS scheme can be found \cite{PARS,CARS}. 
The ARS algorithms are very appealing techniques since they construct a non-parametric proposal to mimic the shape of the target pdf, yielding in general excellent performance (i.e., independent samples from $\bar{\pi}_d$ with a high acceptance rate).

\begin{figure}[htbp]
\centering
\includegraphics[height=5.5cm]{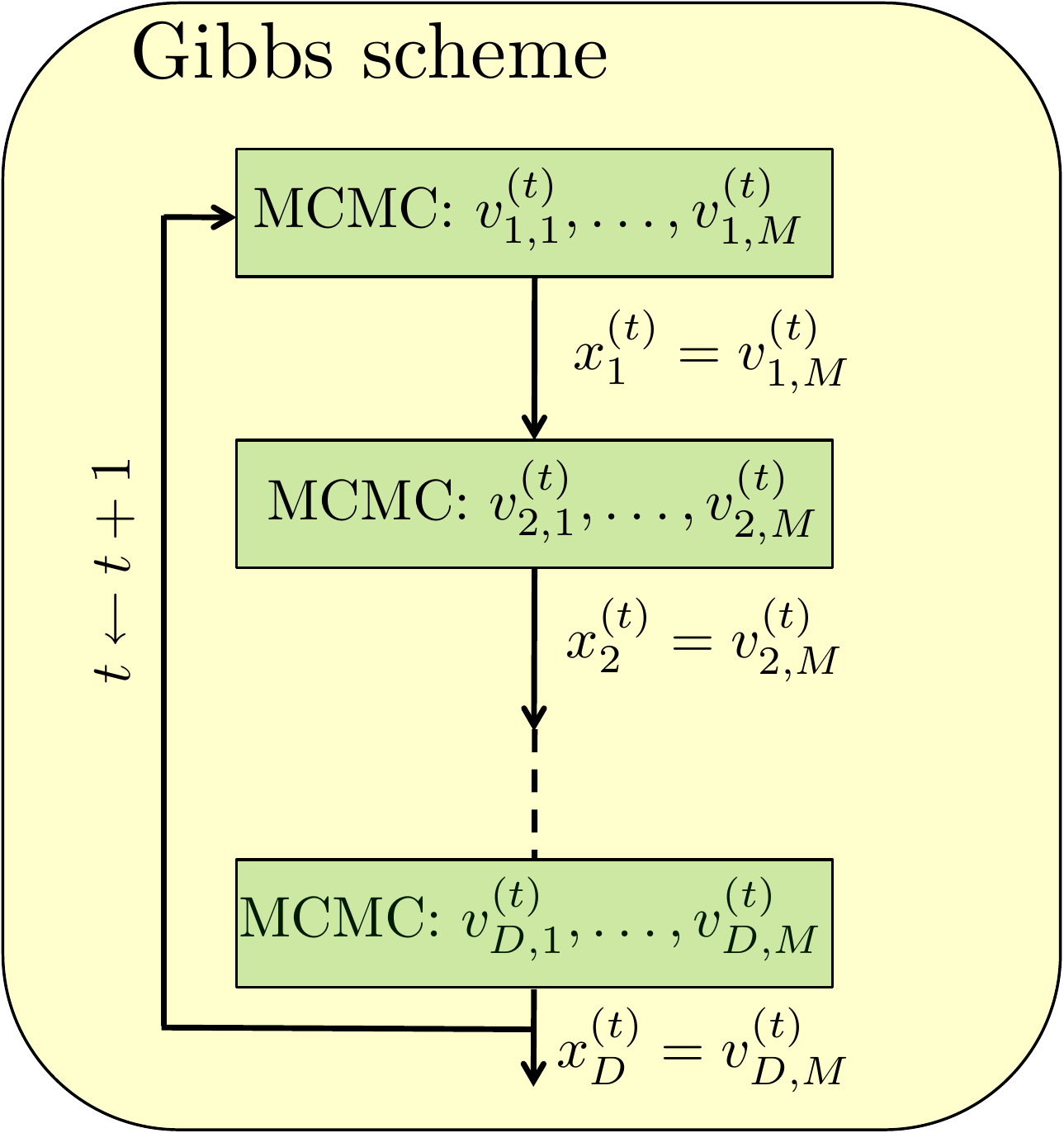} 
\caption{Graphical representation of a generic MCMC-within-Gibbs scheme, where $M$ steps of the internal MCMC algorithm are applied for each full-conditional pdf (see Algorithm~\ref{tab:MCMC_GIBBS}). Different internal MCMC methods have been proposed in literature.
 }
\label{FigMCMCGibbs}
\end{figure}

 However, the range of application of the ARS samplers is limited to some specific classes of densities. Thus, in general, other approaches are required. For instance in \cite{Koch07} an approximated strategy is used, considering the application of the importance sampling (IS) scheme within the Gibbs sampler. A more common approach is to apply an additional MCMC sampler to draw samples from $\bar{\pi}_d$~\cite{Gelfand93}. Therefore, in many practical scenarios, we have an MCMC (e.g., an MH sampler) inside another MCMC scheme (i.e., the Gibbs sampler) as shown in  Figures ~\ref{FigMonteCarloGibbs}-\ref{FigMCMCGibbs}. In the so-called {\it MH-within-Gibbs} approach\footnote{Sometimes MH-within-Gibbs is also referred as to the Single Component MH algorithm~\cite{HaarioCW} or the  Componentwise MH algorithm~\cite{Levine05}.}, only one MH step is often performed within each Gibbs iteration to draw samples from each full-conditional. This hybrid approach preserves the ergodicity of the Gibbs sampler~\cite[Chapter 10]{Robert04}, and provides good performance in many cases. However, several authors have noted that using a single MH step for the internal MCMC is not always the best solution in terms of performance, c.f.~\cite{Brewer93}. 

Using a larger number of iterations of the MH algorithm within-Gibbs can improve the performance~\cite{Muller91,Gelfand93,Fox12}. This is the scenario graphically represented in Figure~\ref{FigMCMCGibbs}. Moreover, different more sophisticated MCMC algorithms to be applied within-Gibbs have been proposed~\cite{Gilks95,Cai08,HaarioCW,ritter1992griddyGibbs}. 
%They provide a sequence of samples with smaller correlation w.r.t. an non-adaptive algorithm, i.e., a scenario more similar to the use of ARS (which produces independent samples)~\cite{}. Several more sophisticated MH schemes for the application ``within-Gibbs'' (adaptive or not) have been proposed in the recent literature~\cite{Cai08,Gilks95,HaarioCW,MartinoA2RMS,FUSS,Levine05,Lucka16,Meyer08,Shao13,Zhang16}. 
Some of these techniques employ  an automatic construction of the proposal density tailored to the specific full-conditional \cite{ritter1992griddyGibbs,Shao13,FUSS}. Other methods use an adaptive parametric proposal pdf~\cite{HaarioCW,Levine05}, while other ones  employ  adaptive non-parametric proposals~\cite{Gilks95,Meyer08,MartinoA2RMS,Sticky13}, in the same fashion of the ARS schemes. It is important to remark here that performing more steps of an adaptive MH method within a Gibbs sampler can provide better results than a longer Gibbs chain applying only one step of a standard MH method~\cite{Gilks95}. Algorithm~\ref{tab:MCMC_GIBBS} describes a generic MCMC-within-Gibbs sampler considering $M$ steps of the  internal MCMC at each Gibbs iteration.
 
While these algorithms are specifically designed to be applied ``within-Gibbs'' and provide very good performance, they still require an increase in the computational cost that is not completely exploited: several samples are drawn from the full-conditionals and used to adapt the proposal pdf, but only a subset of them is employed within the resulting Gibbs estimator. 
%%%%%This chosen sample is also considered in order to  continue with the iterations of the Gibbs sampler. 
%
In this work, we show how they can be incorporated within the corresponding Gibbs estimator to improve performance without jeopardizing its consistency. 
%As also confirmed by the numerical results, this approach is convenient in terms of accuracy and computational efficiency. 
%
In the following, we show the relationships between the standard Gibbs scheme and the chain rule. Then, we describe an alternative formulation of the chain rule useful for introducing the novel Gibbs approach described in Section~\ref{NovelSect}. 
%Furthermore, in general, the ARS methods and the MCMC algorithms which uses a non-parametric proposal pdf~\cite{Gilks95,MartinoA2RMS,FUSS,Meyer08} have also an important benefit: they are able to provide easily a suitable estimation of the normalizing constant of each full-conditional (this is not trivial using MCMC outputs~\cite{Chib01}). %In this work, we also show how to use this information in order to provide a consistent estimator of the marginal likelihood $Z({\bf y})$, related to the joint posterior target distribution.

\begin{algorithm}[t!]
\caption{Generic MCMC-within-SG sampler.\label{tab:MCMC_GIBBS}}
  \begin{algorithmic}[1]
	\STATE{Choose $[x_{1}^{(0)},\ldots,x_{D}^{(0)}]$.}
	\FOR{$t=1,\ldots,T$}
		\FOR{$d=1,\ldots,D$}
			\STATE Perform $M$ steps of an MCMC algorithm with initial state $v_{d,0}^{(t)}=x_d^{(t-1)}$, and target pdf $\bar{\pi}_d(x_d|x_{1:d-1}^{(t)},x_{d+1:D}^{(t-1)})$, yielding the sequence of samples $v_{d,1}^{(t)},\ldots, v_{d,M}^{(t)}$. 
			\STATE Set $x_d^{(t)}=v_{d,M}^{(t)}$. %or alternatively $x_d^{(t)}=v_{d,j^*}^{(t)}$ where $j^*\sim \mathcal{U}(1,\ldots,M)$.
		\ENDFOR
			\STATE Set ${\bf x}^{(t)}=x_{1:D}^{(t)}=[x_1^{(t)},x_2^{(t)},\ldots,x_D^{(t)}]$.
	\ENDFOR
%	\STATE Return $\{{\bf x}^{(t)}\}$ for $t=1,\ldots,T$.
	\STATE Return $\{{\bf x}^{(1)},\ldots,{\bf x}^{(T)}\}$ %for $t=1,\ldots,T$.
\end{algorithmic}
\end{algorithm}

%%%%%%%%%%%%%%%%%%%%%%% %%%%%%%%%%%%%%%%%%%%%%%%%%%
\subsection{Chain rule and the connection with Gibbs sampling}
 %%%%%%%%%%%%%%%%%%%%%%% %% %%%%%%%%%%%%%%%%%%%%%%%%% 
% \cmag{[Se puede considerar como una version MUY especial del Gibbs, donde tienes una marginal y el resto son condicionales respecto a las anteriores... Se que no es un Gibbs, pero por darle coherencia]}
 \label{EstoAltChainRule0}
Let us highlight an important consideration for the derivation of the novel Gibbs approach we will introduce in the following section. For the sake of simplicity, let us consider a bivariate target pdf that can be factorized according to the chain rule, 
\begin{eqnarray*}
{\bar \pi}(x_1,x_2)&=&{\bar \pi}_2(x_2|x_1)p_1(x_1) \\
&=&{\bar \pi}_1(x_1|x_2)p_2(x_2),  
\end{eqnarray*}
where we have denoted with $p_1$, $p_2$, the marginal pdfs of $x_1$ and $\bar\pi_2$, $\bar\pi_1$, are the conditional pdfs. Let us consider the first equality.
Clearly, if we are able to draw from the marginal pdf $p_1(x_1)$ and from the conditional pdf ${\bar \pi}_2(x_2|x_1)$, we can draw samples from ${\bar \pi}(x_1,x_2)$ following the chain rule procedure in Algorithm~\ref{alg:gibbs2}. Note that, consequently, the $T$ independent random vectors  $[x_1^{(t)},x_2^{(t)}]$, with $t=1,\ldots,T$, are all distributed as ${\bar \pi}(x_1,x_2)$.
%\newline
%\newline
%For $t=1,\ldots,T$
%\begin{itemize}
%\item Draw $x_1^{(t)}\sim p_1(x_1)$ and $x_2^{(t)}\sim\bar{\pi}_2(x_2|x_1^{(t)})$.
%\end{itemize}

\begin{algorithm}[h!]
\caption{Chain rule method\label{alg:gibbs2}}
  \begin{algorithmic}[1]
	\FOR{$t=1,\ldots,T$}
		\STATE{Draw $x_1^{(t)}\sim p_1(x_1)$ and $x_2^{(t)}\sim\bar{\pi}_2(x_2|x_1^{(t)})$}
	\ENDFOR
\end{algorithmic}
\end{algorithm}

%%%%%%%%%%%%%%%%%%%%%%% %%%%%%%%%%%%%%%%%%%%%%%%%%%
\subsubsection{Standard Gibbs sampler as the chain rule}
 %%%%%%%%%%%%%%%%%%%%%%% %% %%%%%%%%%%%%%%%%%%%%%% %% %%%%%%
Let us consider again the previous bivariate case where the target pdf is factorized as ${\bar \pi}({\bf x})={\bar \pi}(x_1,x_2)$. The standard Gibbs sampler in this bivariate case consists of the steps in Algorithm~\ref{alg:gibbs3}.
After the burn-in period, the chain converges to the target pdf, i.e., ${\bf x}^{(t)} \sim {\bar \pi}({\bf x})$. Therefore, recalling that ${\bar \pi}(x_1,x_2)={\bar \pi}_2(x_2|x_1)p_1(x_1)={\bar \pi}_1(x_1|x_2)p_2(x_2)$ for $t \geq t_b$, each component of the vector ${\bf x}^{(t)}=[x_1^{(t)},x_2^{(t)}]$ is distributed as the corresponding marginal pdf, i.e., $x_1^{(t)} \sim p_1(x_1)$ and $x_2^{(t)} \sim p_2(x_2)$. Therefore, after $t_b$ iterations, the standard Gibbs sampler can be interpreted as the application of the chain rule procedure in Algorithm~\ref{alg:gibbs2}. Namely, for $t \geq t_b$, Algorithm~\ref{alg:gibbs3} is equivalent to generate $x_1^{(t)}\sim p_1(x_1)$, and then draw $x_2^{(t)}\sim\bar{\pi}_1(x_2|x_1^{(t)})$.

\begin{algorithm}[h!]
\caption{The standard Gibbs sampler for a bivariate target pdf. \label{alg:gibbs3}}
  \begin{algorithmic}[1]
	\FOR{$t=1,\ldots,T$}
		\STATE Draw $x_2^{(t)}\sim\bar{\pi}_2(x_2|x_1^{(t-1)})$.
		\STATE Draw $x_1^{(t)}\sim\bar{\pi}_1(x_1|x_2^{(t)})$.
		\STATE Set ${\bf x}^{(t)}=[x_1^{(t)},x_2^{(t)}]$.
	\ENDFOR
\end{algorithmic}
\end{algorithm}
%\begin{framed}
%\begin{itemize}
%\item For $t=1,\ldots,T$:
%\begin{enumerate}
%\item Draw $x_2^{(t)}\sim\bar{\pi}_1(x_2|x_1^{(t-1)})$.
%\item Draw $x_1^{(t)}\sim\bar{\pi}_2(x_1|x_2^{(t)})$.
%\item Set ${\bf x}^{(t)}=[x_1^{(t)},x_2^{(t)}]$.
%\end{enumerate}
%\end{itemize}
%\end{framed}

%\begin{algorithm}[h!]
%\caption{The Gibbs sampler as the chain rule: equivalent formulation \label{alg:gibbs4}}
%  \begin{algorithmic}[1]
%	\FOR{$t\geq t_b$}
%		\STATE Generate  $x_1^{(t)}\sim p_1(x_1)$, and then draw $x_2^{(t)}\sim\bar{\pi}_1(x_2|x_1^{(t)})$
%		\STATE Set ${\bf x}^{(t)}=[x_1^{(t)},x_2^{(t)}]$
%	\ENDFOR
%\end{algorithmic}
%\end{algorithm}
%\begin{framed}
%\begin{itemize}
%\item For $t\geq t_b$, repeat:
%\begin{enumerate}
%\item Generate  $x_1^{(t)}\sim p_1(x_1)$, and then draw $x_2^{(t)}\sim\bar{\pi}_1(x_2|x_1^{(t)})$ .
%\item Set ${\bf x}^{(t)}=[x_1^{(t)},x_2^{(t)}]$.
%\end{enumerate}
%\end{itemize}
%\end{framed}

%%%%%%%%%%%%%%%%%%%%%%% %%%%%%%%%%%%%%%%%%%%%%%%%%%
\subsubsection{Alternative chain rule procedure}
 %%%%%%%%%%%%%%%%%%%%%%% %% %%%%%%%%%%%%%%%%%%%%%%%%% 
 \label{EstoAltChainRule}
An alternative procedure is shown in Algorithm~\ref{alg:chain1}. This chain rule draws $M$ samples from the full conditional $\bar{\pi}_2(x_2|x_1)$ at each $t$-th iteration, and generates samples from the joint pdf $\bar{\pi}(x_1,x_2)$. 

\begin{algorithm}[h!]
\caption{An alternative chain rule procedure. \label{alg:chain1}}
  \begin{algorithmic}[1]
	\FOR{$t=1,\ldots,T$}
		\STATE Draw $x_1^{(t)}\sim p_1(x_1)$.
		\STATE Draw $x_{2,m}^{(t)}\sim\bar{\pi}_2(x_2|x_1^{(t)})$, with $m=1,\ldots.M$.
	\ENDFOR
\end{algorithmic}
\end{algorithm}

\begin{figure}[t!]
\begin{center}
\begin{tabular}{ccc}
(a) Standard sampling & & (b) Alternative sampling\\
\includegraphics[height=3cm]{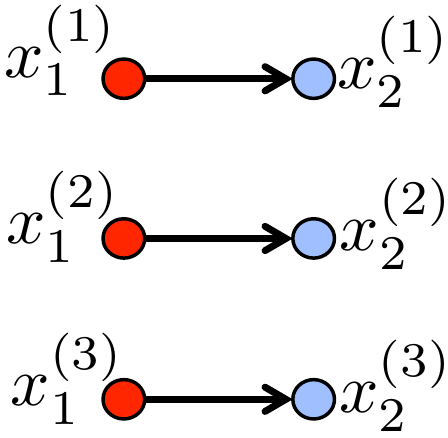} & & \includegraphics[height=3cm]{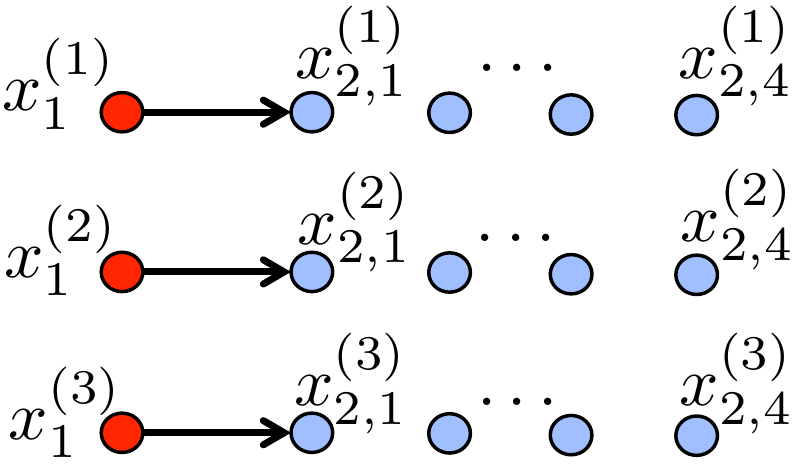}
\end{tabular}
\caption{Graphical representation of the (a) standard chain rule sampling ($M=1$), and (b) the alternative chain rule sampling ($M=4$). In both cases, $N=3$. The total number of drawn vectors $[x_1^{(t)},x_{2,m}^{(t)}] \sim {\bar \pi}(x_1,x_2)={\bar \pi}_2(x_2|x_1)p_1(x_1)$ is $NM=3$ and $NM=12$, respectively.}
\label{FigChainRule}
\end{center}
 \end{figure}

Note that all the $TM$ vectors,  $[x_{1}^{(t)},x_{2,m}^{(t)}]$, with $t=1,\ldots,T$ and $m=1,\ldots,M$, are samples from ${\bar \pi}(x_1,x_2)$. This scheme is valid and, in some cases, can present some benefits w.r.t. the traditional scheme in terms of performance, depending on  some characteristics contained in the joint pdf ${\bar \pi}(x_1,x_2)$. For instance, the correlation between variables $x_1$ and $x_2$, and the variances of the marginal pdfs $p_1(x_1)$ and $p_2(x_2)$. Figure~\ref{FigChainRule} shows the graphical representation of the standard chain rule sampling scheme (with $T=3$ and $M=1$), and the alternative chain rule sampling procedure described before (with $T=3$, $M=4$).

%\newline
%\newline
%For $t=1,\ldots,T$
%\begin{enumerate}
%\item Draw $x_1^{(t)}\sim p_1(x_1)$.
%%\item Draw $M$ (conditionally) independent samples, $x_2^{(M(n-1)+m)}\sim\bar{\pi}_2(x_2|x_1^{(n)})$, with $m=1,\ldots.M$.
%\item Draw $x_{2,m}^{(t)}\sim\bar{\pi}_2(x_2|x_1^{(t)})$, with $m=1,\ldots.M$.
%\end{enumerate}

\begin{figure}[htbp]
\begin{center}
\centerline{
\includegraphics[width=0.6\textwidth]{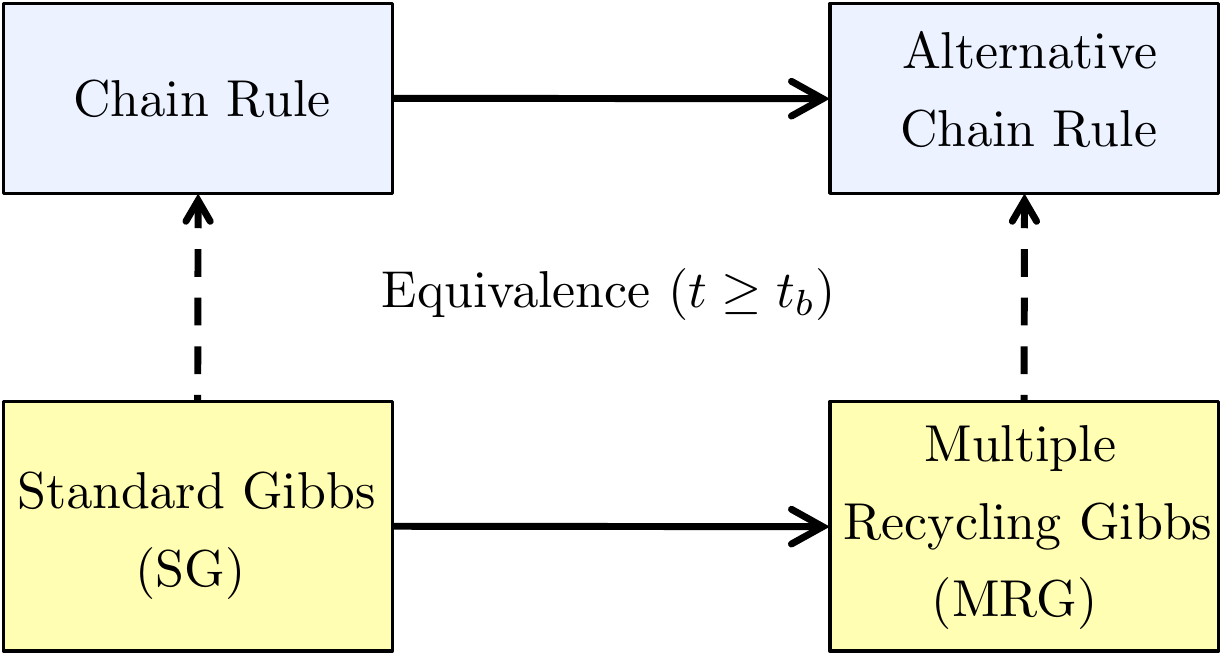}%\hspace{1cm}
}
\caption{Graphical representation of the relationships between chain rules and Gibbs schemes.} %{  In the chain rule method the $M$ samples are independently generated, whereas in the SG sampler they are correlated (the correlation does not jeopardize the consistency).}  }
\label{FigTeo}
\end{center}
\end{figure}

At this point, a natural question arises: is it possible to design a Gibbs sampling scheme equivalent to the alternative chain rule scheme described before? In the next section, we introduce the Multiple Recycling Gibbs Sampler (MRG), which corresponds to the alternative chain rule procedure, as summarized in Fig.~\ref{FigTeo}.

%{ falta enlar con la alternative formulation of the chain rule}
%%%using some MCMC procedure with invariant pdf

%%%%%%%%%%%%%%%%%%%%%%%%%%%%%%%%%%%%%%%%%%%%%%%%%%%%%%%%%%
\section{The Recycling Gibbs sampler}
%%%%%%%%%%%%%%%%%%%%%%%%%%%%%%%%%%%%%%%%%%%%%%%%%%%%%%%%%%
\label{NovelSect}
The previous considerations suggest that we can benefit from some previous intermediate points produced in the Gibbs procedure. More specifically, let us consider the following {\it Trivial Recycling Gibbs} (TRG) procedure in Algorithm~\ref{alg:trg1}.
\begin{algorithm}[h!]
\caption{Trivial Recycling Gibbs (TRG) procedure. \label{alg:trg1}}
  \begin{algorithmic}[1]
	\FOR{$t=1,\ldots,T$}
	\FOR{$d=1,\ldots,D$}
		\STATE Draw $x_d^{(t)}\sim\bar{\pi}_d(x_d|x_{1:d-1}^{(t)},x_{d+1:D}^{(t-1)})$.
		\STATE Set ${\bf x}_d^{(t)}=[x_{1:d-1}^{(t)},x_d^{(t)},x_{d+1:D}^{(t-1)}]=[x_{1:d}^{(t)},x_{d+1:D}^{(t-1)}]$.
	\ENDFOR
	\ENDFOR
	\RETURN{Return $\{{\bf x}_{d}^{(t)}\}$ for all $d$ and $t$.}
\end{algorithmic}
\end{algorithm}

\begin{figure}[htbp]
\begin{center}
\setlength{\tabcolsep}{0pt}
\begin{tabular}{ccc}
%(a) Gibbs sampler & (b) Recycling Gibbs sampler & (c) Multiple Recycling Gibbs sampler\\
(a) SG & (b) TRG & (c) MRG\\
\includegraphics[height=3.8cm]{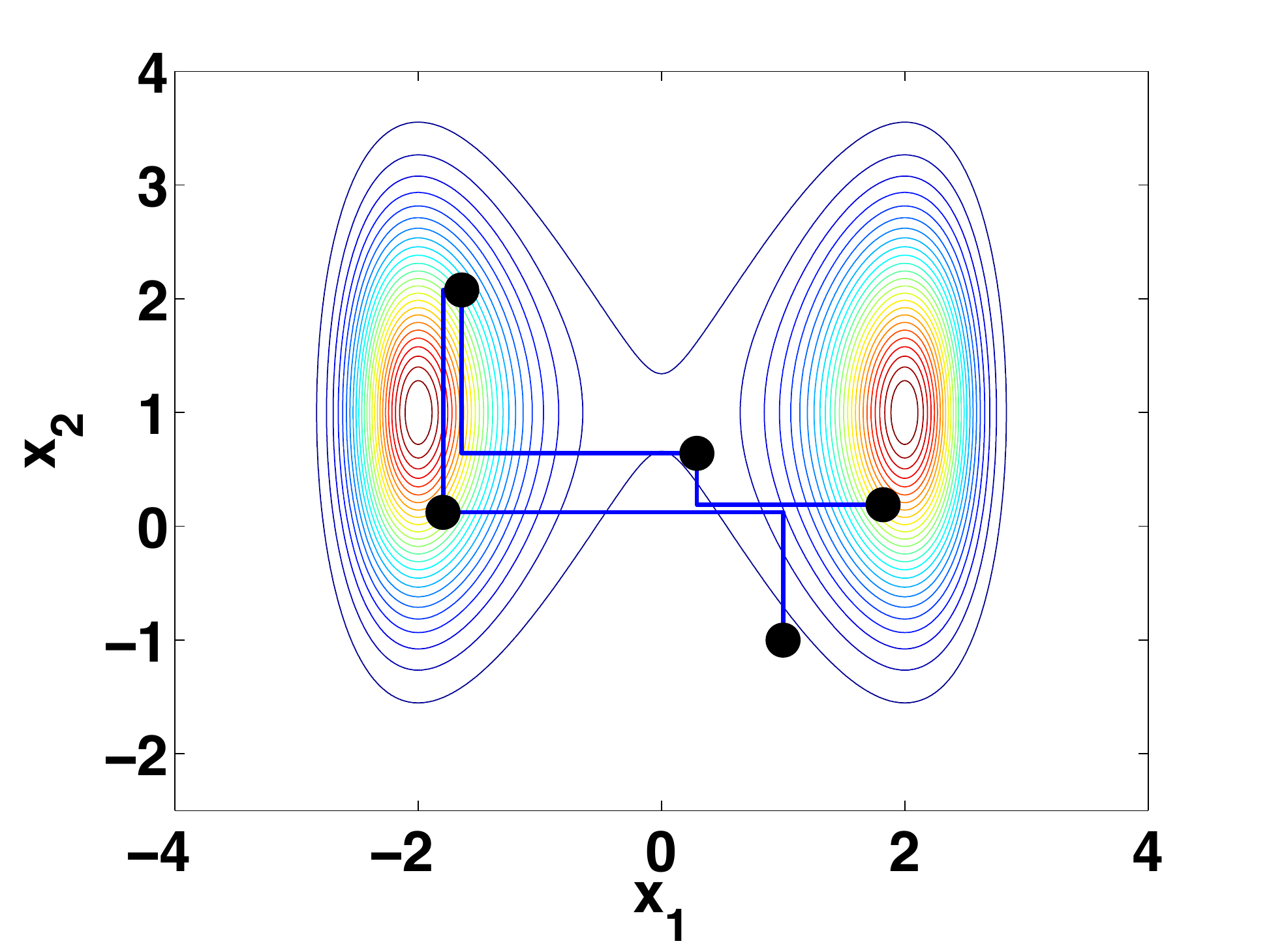} & 
\includegraphics[height=3.8cm]{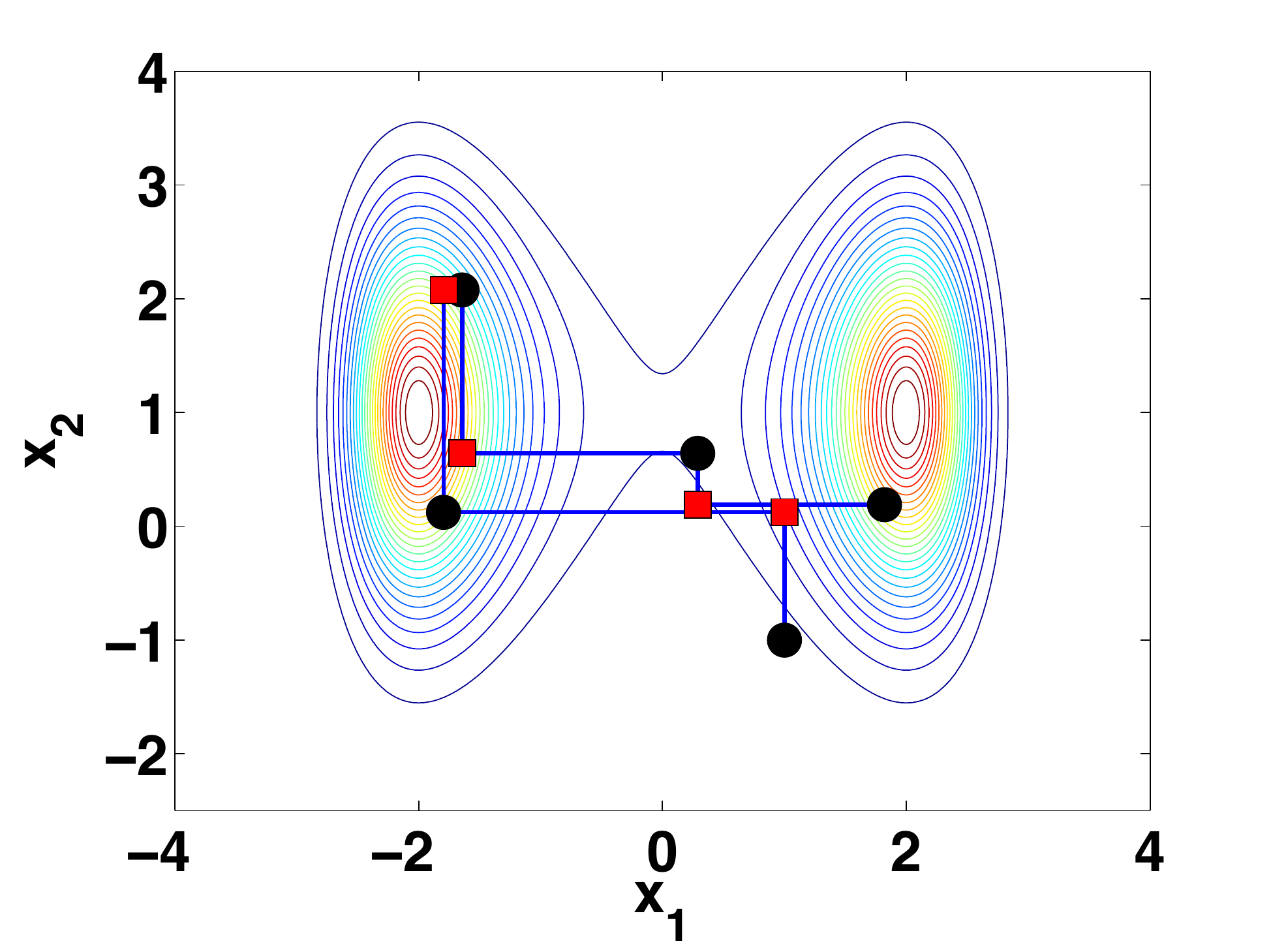} & 
\includegraphics[height=3.8cm]{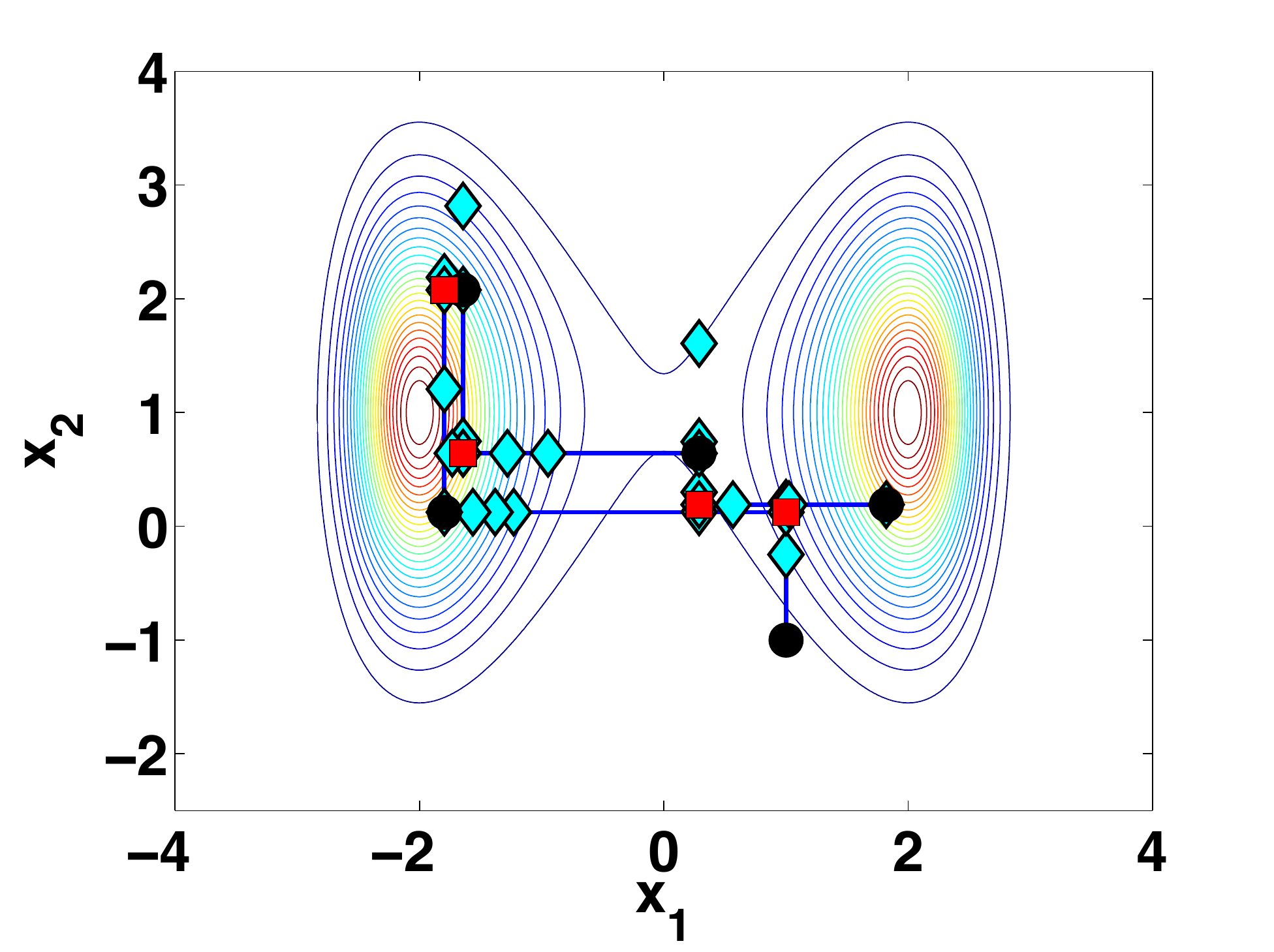} 
\end{tabular}
\caption{We consider $T=4$ iterations of a Gibbs sampler and $M=5$ iterations of the MH for drawing from each full-conditional pdfs. {\bf (a)} With the circles we denote  the $T+1$ points (considering the starting point) used in the standard Gibbs estimators.  {\bf (b)} The vectors (denoted with circles and squares) used in the TRG estimators. {\bf (c)} The vectors (denoted with circles, squares and diamonds) used in the MRG estimators.}
\label{FigTodo}
\end{center}
\end{figure}

The procedure generates $DT$ samples ${\bf x}_d^{(t)}$, with $d=1,\ldots,D$ and $t=1,\ldots,T$, shown in Figure~\ref{FigTodo}(b) with circles and squares. Note that if we consider only the subset of generated vectors 
$$
{\bf x}_{D}^{(t)}, \quad t=1,\ldots, T,
$$
by setting $d=D$, we obtain the outputs of the standard Gibbs (SG) sampler approach in Algorithm \ref{alg:gibbs}. Namely, the samples generated by a SG procedure can be obtained by subsampling the samples obtained by the proposed RG. 
% the set of the samples obtained with the standard Gibbs (SG) method, considering the same generated components $x_d^{(t)}$ for all $d$ and $t$.  First of all, clearly we have 
% $\mathcal{S}_{SG} \subseteq \mathcal{S}_{RG}$,
% since 
% $$
% {\bf x}_{SG}^{(t)}={\bf x}_{RG}^{(Dt)},
% $$
% i.e., the samples obtained by a SG can be obtained by a subsampling of the samples obtained by RG.
Figure~\ref{FigTodo}(a) depicts with circles $T+1$ vectors (considering also the starting point) corresponding to a run of SG with $T=4$. Figure~\ref{FigTodo}(b) shows with squares the additional points used in TRG. 

Let us consider the estimation by SG and TRG of a generic moment, i.e.,  given a function $f(x_d)$, of $d$-th marginal density, i.e., $p_d(x_d)=\int_{\mathbb{R}^{D-1}} \bar{\pi}({\bf x}) d{\bf x}_{\neg d} $.
After a closer inspection,  we note that both estimators corresponding to the SG and TRG coincide:   
\begin{eqnarray}
\int_{\mathbb{R}^{D}} f(x_d) \bar{\pi}({\bf x}) d{\bf x}=\int_{\mathbb{R}} f(x_d) p_d(x_d) dx_d\approx \frac{1}{DT} \sum_{t=1}^{T} \sum_{d=1}^{D} f(x_{d}^{(t)})=\frac{1}{T} \sum_{t=1}^{T} f(x_{D}^{(t)}),  
\end{eqnarray}
where for the last equality we are assuming (for the sake of simplicity) that the $d$-th component is the second variable in the Gibbs scan  and $T=kD$, $k\in \mathbb{N}$. This is due to the fact that, in TRG, each component $x_{d}^{(t)}$ is repeated exactly $D$ times (inside different consecutive samples) and we have $D$ times more samples in TRG than in a standard SG. Hence, in such situation, there are no apparent advantages of using TRG w.r.t. a SG approach. Namely, TRG and SG are equivalent schemes in the approximation of the marginal densities (we remark that the expression above is valid only for marginal moments). The advantages of a RG strategy appear clear when more than one sample is drawn from the full-conditional, $M>1$, as discussed below. 

%%%

Based on the previous considerations, we design the {\it  Multiple Recycling Gibbs} (MRG) sampler which draws $M>1$ samples from each full conditional pdf, as shown in Algorithm~\ref{tab:MRG}.  Figure~\ref{FigTodo}(c) shows all the samples (denoted with circles, squares and diamonds) used in the MRG estimators. Thus, given a specific function $f({\bf x})$ in the integral in Eq.~\eqref{MainInt}, the MRG estimator is eventually formed by $TDM$ samples, without removing any burn-in period,
\begin{equation}
\widehat{I}_T=\frac{1}{TDM} \sum_{t=1}^{T} \sum_{d=1}^{D}  \sum_{m=1}^{M}  f({\bf x}_{d,m}^{(t)}). 
\end{equation}
Observe that in order to go forward to sampling from the next full-conditional, we only consider the last generated component, i.e., $z_d^{(t)}=x_{d,M}^{(t)}$. However, an alternative to step~\ref{ThisStep} of Algorithm~\ref{tab:MRG} is: (a) draw $j\sim \mathcal{U}(1,\ldots,M)$ and (b) set $z_d^{(t)}=x_{d,j}^{(t)}$. Note that choosing the last sample $x_{d,M}^{(t)}$ is more convenient for an MCMC-within-MRG scheme.

\begin{algorithm}[h!]
\caption{Multiple Recycling Gibbs (MRG) sampler. \label{tab:MRG}}
  \begin{algorithmic}[1]
	\STATE Choose a starting point $[z_1^{(0)},\ldots,z_D^{(0)}]$.
	\FOR{$t=1,\ldots,T$}
	\FOR{$d=1,\ldots,D$}
	\FOR{$m=1,\ldots,M$}
		\STATE Draw $x_{d,m}^{(t)}\sim\bar{\pi}_d(x_d|z_{1:d-1}^{(t)},z_{d+1:D}^{(t-1)})$.
		\STATE Set ${\bf x}_{d,m}^{(t)}=[z_{1:d-1}^{(t)}, x_{d,m}^{(t)},z_{d+1:D}^{(t-1)}]$.
	\ENDFOR
	\STATE \label{ThisStep} Set $z_d^{(t)}=x_{d,M}^{(t)}$.
	\ENDFOR
	\ENDFOR
\RETURN{Return $\{{\bf x}_{d,m}^{(t)}\}$ for all $d$, $m$ and $t$.}
\end{algorithmic}
\end{algorithm}

As shown in Figure~\ref{FigTeo}, MRG is equivalent to the alternative chain rule scheme described in the previous section, so that the consistency of the MRG estimators is guaranteed. The ergodicity of the generated chain is also ensured since the dynamics of the MRG scheme is identical to the dynamics of the SG sampler (they differ in the construction of final estimators). Note that with $M=1$, we go back to the TRG scheme. 

The MRG approach is convenient in terms of accuracy and computational efficiency, as also confirmed by the numerical results in Section~\ref{SIMU}. MRG is particularly advisable if an adaptive MCMC is employed to draw from the full-conditional pdfs, i.e., when several MCMC steps are performed for sampling from each full-conditional and adapting the proposal. We can use all the sequence of samples generated by the internal MCMC algorithm in the resulting estimator.  %Algorithm~\ref{tab:MRG} describes the MRG scheme when is possible to draw directly from the full-conditional pdfs.  
Algorithm~\ref{tab:MCMC_MRG} shows the detailed steps of an MCMC-within-MRG algorithm, when a direct method for sampling the full-conditionals is not available.

\begin{algorithm}[h!]
\caption{Generic MCMC-within-MRG sampler. \label{tab:MCMC_MRG}}
  \begin{algorithmic}[1]
	\STATE Choose a starting point $[z_1^{(0)},\ldots,z_D^{(0)}]$.
	\FOR{$t=1,\ldots,T$}
	\FOR{$d=1,\ldots,D$}
		\STATE Perform $M$ steps of an MCMC algorithm with target pdf $\bar{\pi}_d(x_d|x_{1:d-1}^{(t)},x_{d+1:D}^{(t-1)})$, yielding the sequence of samples $x_{d,1}^{(t)},\ldots, x_{d,M}^{(t)}$,  with initial state $x_{d,0}^{(t)}=x_d^{(t-1)}$.
\STATE Set ${\bf x}_{d,m}^{(t)}=[z_{1:d-1}^{(t)}, x_{d,m}^{(t)},z_{d+1:D}^{(t-1)}]$, for $m=1,\ldots,M$. %where $j=MD(t-1)+M(d-1)+m$.
\STATE Set $z_d^{(t)}= x_{d,M}^{(t)}$.%\footnote{Alternatively, Draw $j^*\sim \mathcal{U}(1,\ldots,M)$ and set ${\bf z}_d^{(t)}=x_{d,j^*}^{(t)}$.}
%\item Set $x_d^{(t)}={\bf z}_{d,M}^{(t)}$, or alternatively $x_d^{(t)}={\bf z}_{d,j^*}^{(t)}$ where $j^*\sim \mathcal{U}(1,\ldots,M)$.
	\ENDFOR
	\ENDFOR
\RETURN{Return $\{{\bf x}_{d,m}^{(t)}\}$ for all $d$, $m$ and $t$.}
\end{algorithmic}
\end{algorithm}

\begin{figure}[htbp]
\begin{center}
\setlength{\tabcolsep}{0pt}
\begin{tabular}{ccc}
%(a) Gibbs sampler & (b) Recycling Gibbs sampler & (c) Multiple Recycling Gibbs sampler\\
(a) SG & (b) MRG & (c) Histogram - MRG\\
\includegraphics[height=3.8cm]{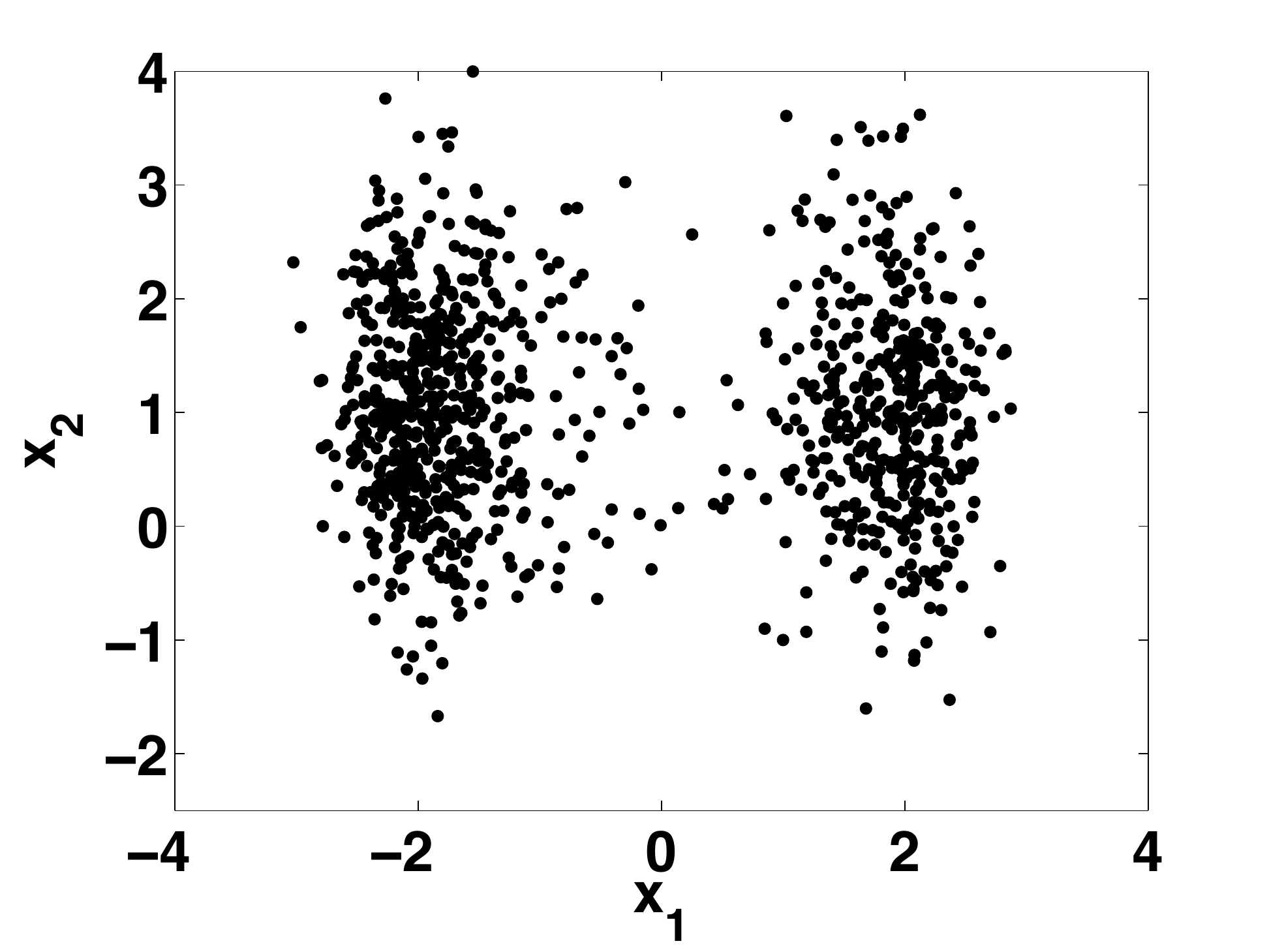} & 
\includegraphics[height=3.8cm]{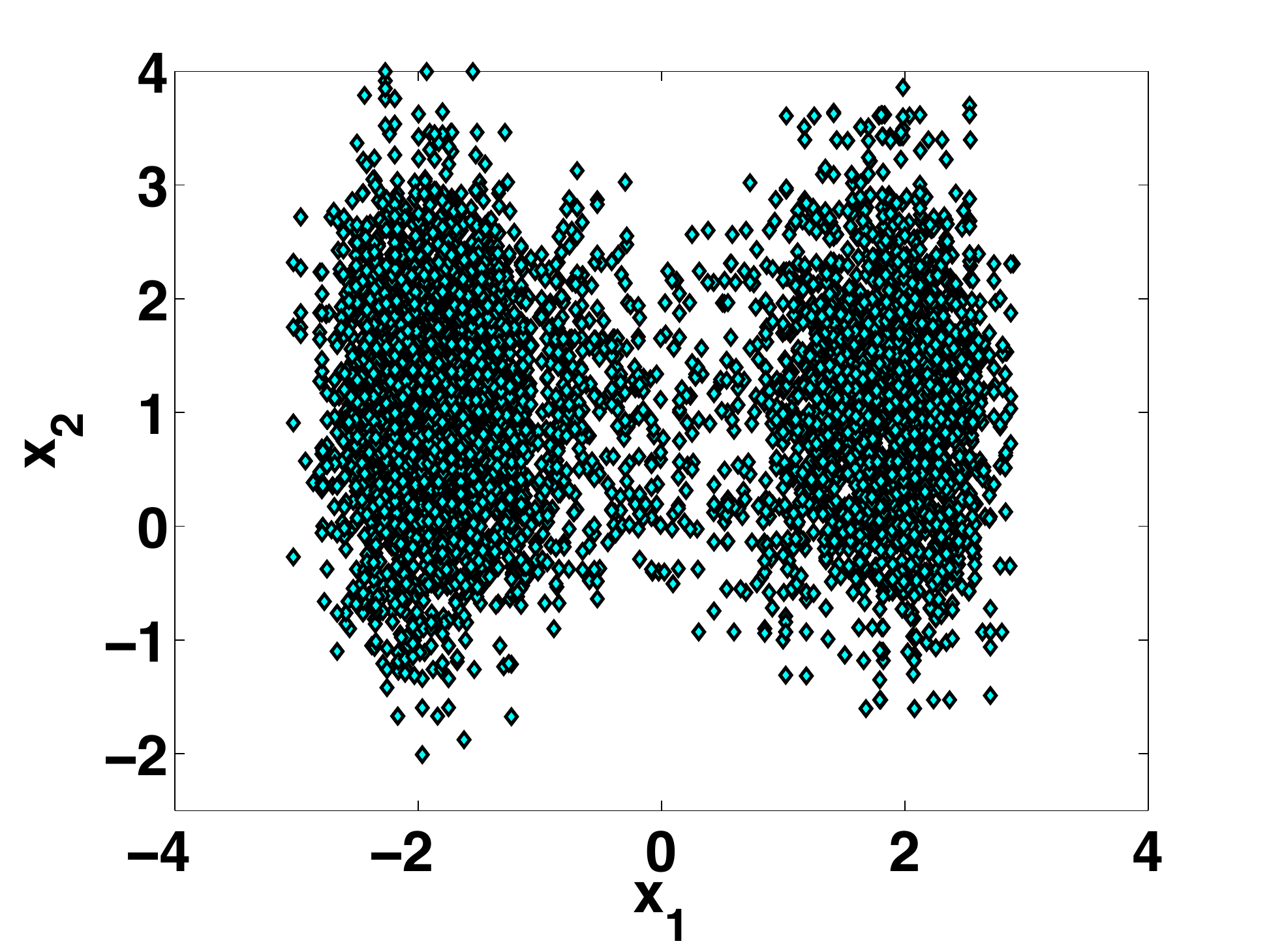} & 
\includegraphics[height=3.8cm]{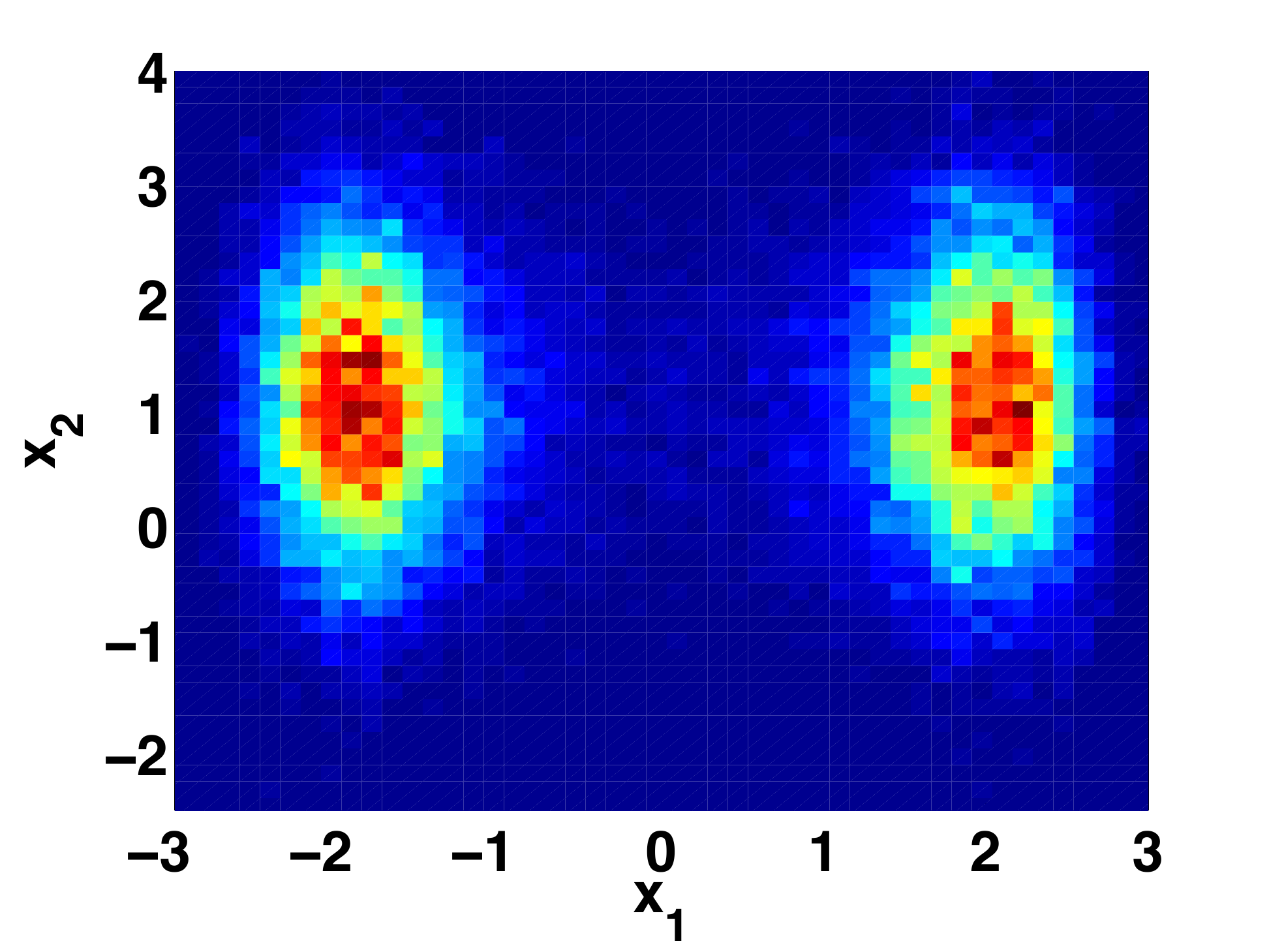} 
\end{tabular}
\caption{{\bf (a)} Outputs of one MH-within-Gibbs run with $T=10^3$ and $M=5$, considering the target with contour plot shown in Fig.~\ref{FigTodo}. {\bf (b)}  Outputs of one MH-within-MRG run with $T=10^3$ and $M=5$. {\bf (c)} Histograms obtained using all the points in Figure (b), i.e., the MRG outputs with  $T=10^4$ and $M=5$.}
\label{FigTodo3}
\end{center}
\end{figure}

Figure~\ref{FigTodo3}(a) depicts the random vectors  obtained with one run of an MH-within-Gibbs procedure, with $T=10^3$ and $M=5$. Figure~\ref{FigTodo3}(b) illustrates all the outputs of the previous run, including all the auxiliary samples generated by the MH algorithm. Hence, these vectors are the samples obtained with a MH-within-MRG approach. The histogram of the samples in Figure~\ref{FigTodo3}(b) is depicted Figure~\ref{FigTodo3}(c). Note that the histogram of the MH-within-MRG samples reproduces adequately the shape of the target pdf shown in Figure~\ref{FigTodo}. This histogram was obtained with one run of MH-within-MRG fixing $T=10^4$ and $M=5$.
 
\iffalse
\begin{table}[!t]
%	\centering
%\small
\caption{Multiple Recycling Gibbs (MRG) sampler. 
}
\begin{tabular}{|p{0.95\columnwidth}|}
\hline
\footnotesize
%\begin{enumerate}
%\item 
\\
- Choose a starting point $[z_1^{(0)},\ldots,z_D^{(0)}]$.
\begin{enumerate}
\item For $t=1,\ldots,T$:
\begin{enumerate}
\item For $d=1,\ldots,D$:
\begin{enumerate}
\item  For $m=1,\ldots,M$:
\begin{enumerate}
\item Draw $x_{d,m}^{(t)}\sim\bar{\pi}_d(x_d|z_{1:d-1}^{(t)},z_{d+1:D}^{(t-1)})$.
\item Set ${\bf x}_{d,m}^{(t)}=[z_{1:d-1}^{(t)}, x_{d,m}^{(t)},z_{d+1:D}^{(t-1)}]$. %where $j=MD(t-1)+M(d-1)+m$.
\end{enumerate}
\item\label{ThisStep} Set $z_d^{(t)}=x_{d,M}^{(t)}$.
\end{enumerate}
\end{enumerate}
\item Return $\{{\bf x}_{d,m}^{(t)}\}$ for all $d$, $m$ and $t$.
\end{enumerate} \\
\hline 
\end{tabular}
\label{tab:MRG}
\end{table}
\fi

 %%%%%%%%%%%%%%%%%%%%%%% %%%%%%%%%%%%%%%%%%%%%%% %%%%%%%%%%%%%%%%%%%%%%%
\section{Experimental Results} \label{SIMU}

This section gives experimental evidence of performance of the proposed scheme. { First of all, we study the efficiency of the proposed scheme in three bi-dimensional toy examples with different target densities: a unimodal Gaussian target presenting linear correlation among the variables, a bimodal target (see Fig. \ref{FigTodo}) and an ``elliptical'' target which presents a strong nonlinear correlation between the variables, as shown in Fig.  \ref{FigSIMUdonut_2}(b).} Then, we show its use in a hyper-parameter estimation problem using Gaussian process (GP) regression~\cite{rasmussen2006gaussian} with a kernel function usually employed for automatic relevance determination (ARD) of the input features, considering different dimension of the inference problem. Results show the advantages of the MRG scheme in all the experiments. Furthermore, we apply MRG in a dependence detection problem using both real and simulated remote sensing data. For the sake of reproducibility, the interested reader may find related source codes in \url{http://isp.uv.es/code/RG.zip}.

{
%%%%%%%%%%%%%%%%%%%%%%%%%%%%%%
\subsection{Experiment 1: A first analysis of the efficiency}
\label{ExFirstAn}
%%%%%%%%%%%%%%%%%%%%%%%%%%%%%%
Let us consider two Gaussian full-conditional densities,
\begin{align}
\label{FullEx0}
	\bar{\pi}_1(x_1|x_2) & \propto \exp\left(-\frac{(x_1-0.5x_2)^2}{2\delta^2}\right), \\
	\bar{\pi}_2(x_2|x_1) & \propto \exp\left(-\frac{(x_2-0.5x_1)^2}{2\delta^2}\right),
\end{align}
with $\delta=1$. The joint target pdf $\bar{\pi}(x_1,x_2)=\mathcal{N}(x_1,x_2|\boldsymbol\mu,\boldsymbol\Sigma)$ is a bivariate Gaussian pdf with mean vector $\boldsymbol\mu = [0,0]^{\top}$ and covariance matrix $\boldsymbol\Sigma = [1.33 \ 066; \ 0.66 \ 1.33]$. Hence, note that $x_1$ and $x_2$ are linearly correlated.
We apply a Gibbs sampler with $T$ iterations to estimate both the mean and the covariance matrix of the joint target pdf. Then, we estimate $5$ values ($2$ for the mean $\boldsymbol\mu$ and $3$ for the covariance matrix $\boldsymbol\Sigma$ ) and takes the average Mean Square Error (MSE).
The results are averaged over $2000$ independent runs.

In this toy example, it is possible to draw directly from the conditional pdfs in Eq. \eqref{FullEx0} since they are both Gaussian densities. Thus, we can show the performance of the two ideal Gibbs schemes: the Ideal SG method (see Alg. \ref{alg:gibbs})  and the Ideal MRG technique (see Alg. \ref{tab:MRG}). Furthermore, we test a standard MH method and an Adaptive MH (AMH) technique \cite{Haario01} within SG and MRG. For both MH and AMH,  we use a Gaussian random walk proposal, $q(x_{d,m}^{(t)}|x_{d,m-1}^{(t)}) \propto \exp\left(-\frac{(x_{d,m}^{(t)}-x_{d,m-1}^{(t)})^2}{2\sigma^2}\right)$, 
for $d \in \{1,2\}$, $1 \le m \le M$ and $1 \le t \le T$.\footnote{{We have also tried the use of a $t$-Student's density as proposal pdf. Clearly, depending on the proposal parameter employed, the MSE values change, in general. However, the comparison among the different Monte Carlo schemes is not affected by the use of a different proposal density.}} We test different values of the scale parameter $\sigma\in\{0.5,1\}$. At each iteration, AMH adapts the value of $\sigma$ as $m$ grows, using the generated samples from the corresponding full-conditional. 

First, we set  $T=1000$ and vary $M$. The results are given in Figure \ref{FigSIMU0_1}(a). Then, we keep fixed $M=20$ and vary $T$, as shown Figure \ref{FigSIMU0_1}(b). In both figures, all the MRG schemes are depicted with solid lines whereas all the SG methods are shown with dashed lines. We can observe that the MRG schemes always provide smaller MSE values. The performance of the MH-within-Gibbs methods depends sensibly on the choice of $\sigma$, showing the importance of using an adaptive technique. Obviously, the MSE of Ideal SG is constant in Figure \ref{FigSIMU0_1}(a) (as function of $M$), and the Ideal MRG give a considerable improvement in both scenarios (varying $M$ or $T$).  Note that, in Figure \ref{FigSIMU0_1}(b), the difference between the Ideal SG and Ideal MRG techniques increases as $T$ grows.
As expected, in both figures we can observe that the MH-within-Gibbs methods need to increase $M$ in order to approach the performance of the corresponding Ideal Gibbs schemes.

In Figure \ref{FigSIMU0_2}(a), we show the MSE of AMH as function of the number of target evaluation per full-conditional $E=MT$.\footnote{{ The  total number of target evaluations for all the tested algorithms is $E=DMT$ ($D=2$ in this case). Since the factor $D$ (dimension of the problem and number of the full-conditionals) is common for all the samplers, we consider $E=MT$.} }  AMH adapts the scale parameter $\sigma$ as $M$ grows (starting with $\sigma=1$).  We consider $M=1$ and $M=10$ and vary $T$ in order to provide the same value of $E$. Clearly, the chain corresponding to AMH-within-Gibbs with $M=10$ is always shorter than the chain of  AMH-within-Gibbs with $M=1$. We can see that  AMH with $M=10$ takes advantage of the adaptation and provides smaller MSE value. Furthermore, with a Matlab implementation, the use of a Gibbs sampler with a greater $M$ and smaller $T$ seems a  faster solution, as shown in Figure \ref{FigSIMU0_2}(b).

  \begin{figure}[h!]
\begin{center}
\centerline{
 \subfigure[MSE as function of $M$ ($T=1000$).]{\includegraphics[width=0.5\textwidth]{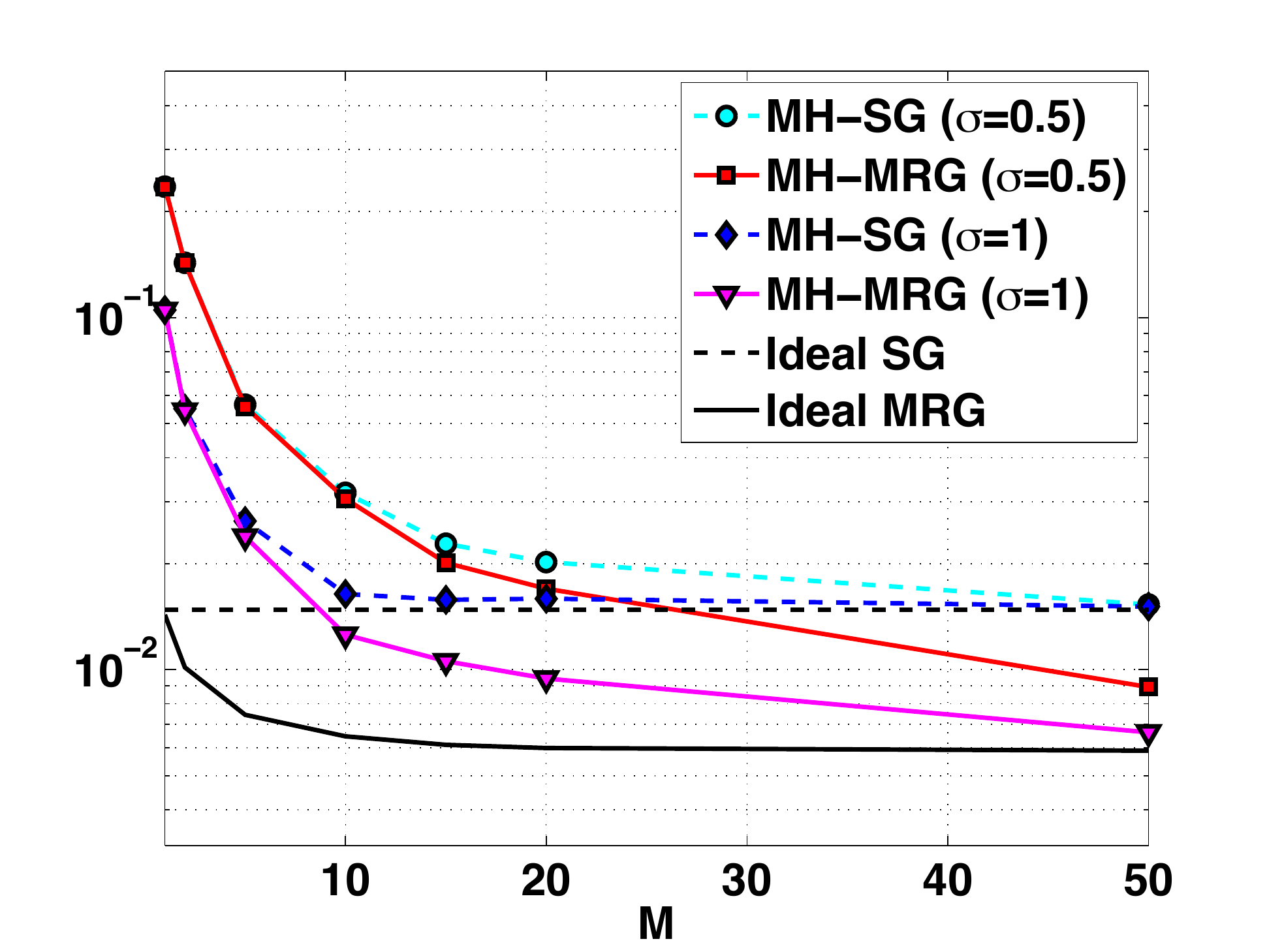}}
  \subfigure[MSE as function of $T$ ($M=20$).]{\includegraphics[width=0.5\textwidth]{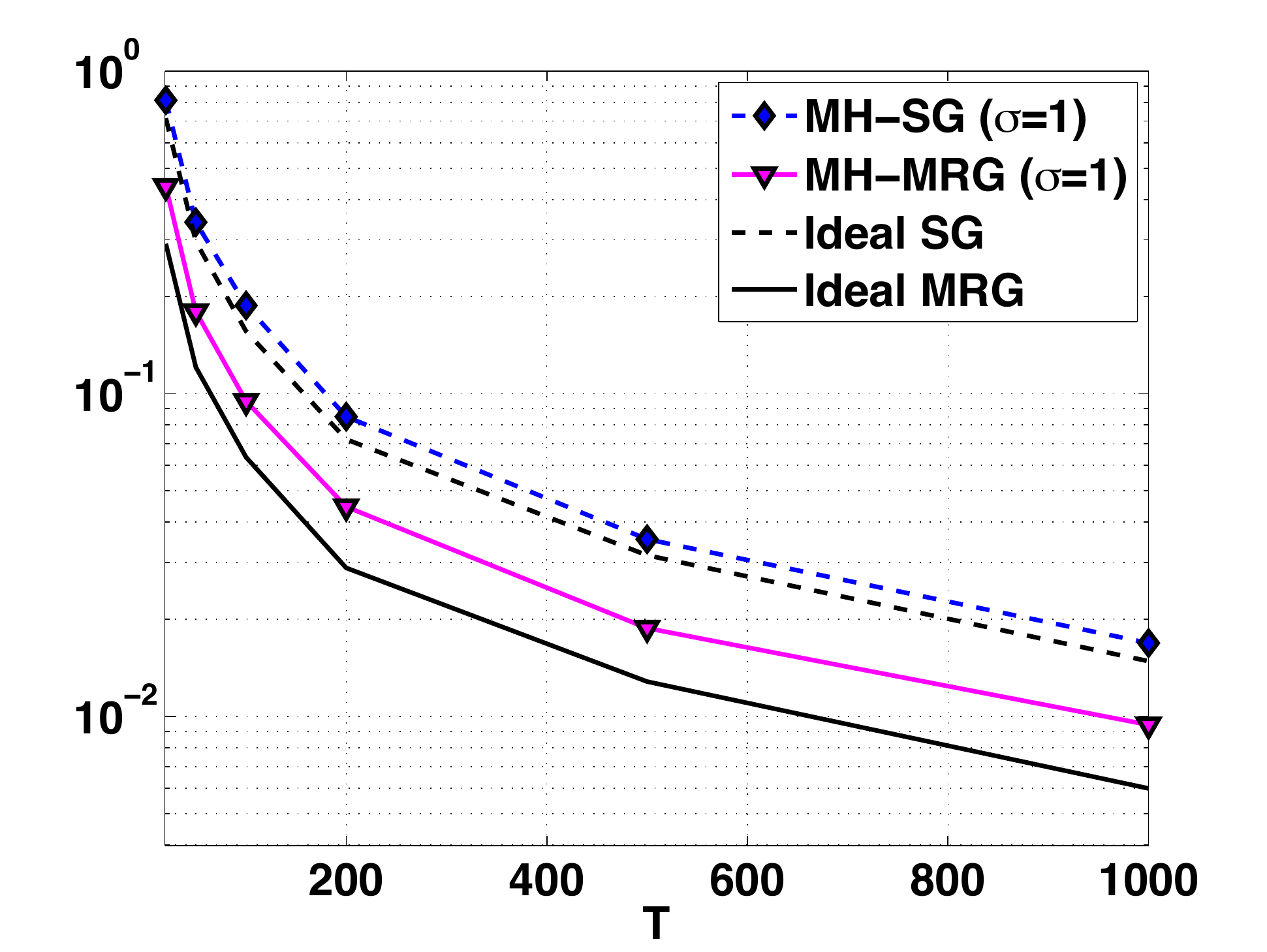}}
%  \subfigure[MSE as function of $M$ with $MT=10^4$ fixed.]{\includegraphics[width=0.5\textwidth]{Fig_MHG_MTfixed.pdf}}
}
\caption{{{\bf Exp. Section \ref{ExFirstAn}-}  {\bf (a)} MSE (log-scale) as function of $M$ for Ideal Gibbs techniques and different MCMC-within-Gibbs schemes (we fix $T=1000$). {\bf (b)} MSE (log-scale) as function of $T$ keeping fixed $M=20$. Note that in both Figures {\bf (a)} and {\bf (b)}, all the MRG methods are depicted with solid lines whereas all the SG schemes with dashed lines.}}
\label{FigSIMU0_1}
\end{center}
\end{figure}

  \begin{figure}[h!]
\begin{center}
\centerline{
 \subfigure[MSE as function of $E=MT$.]{\includegraphics[width=0.5\textwidth]{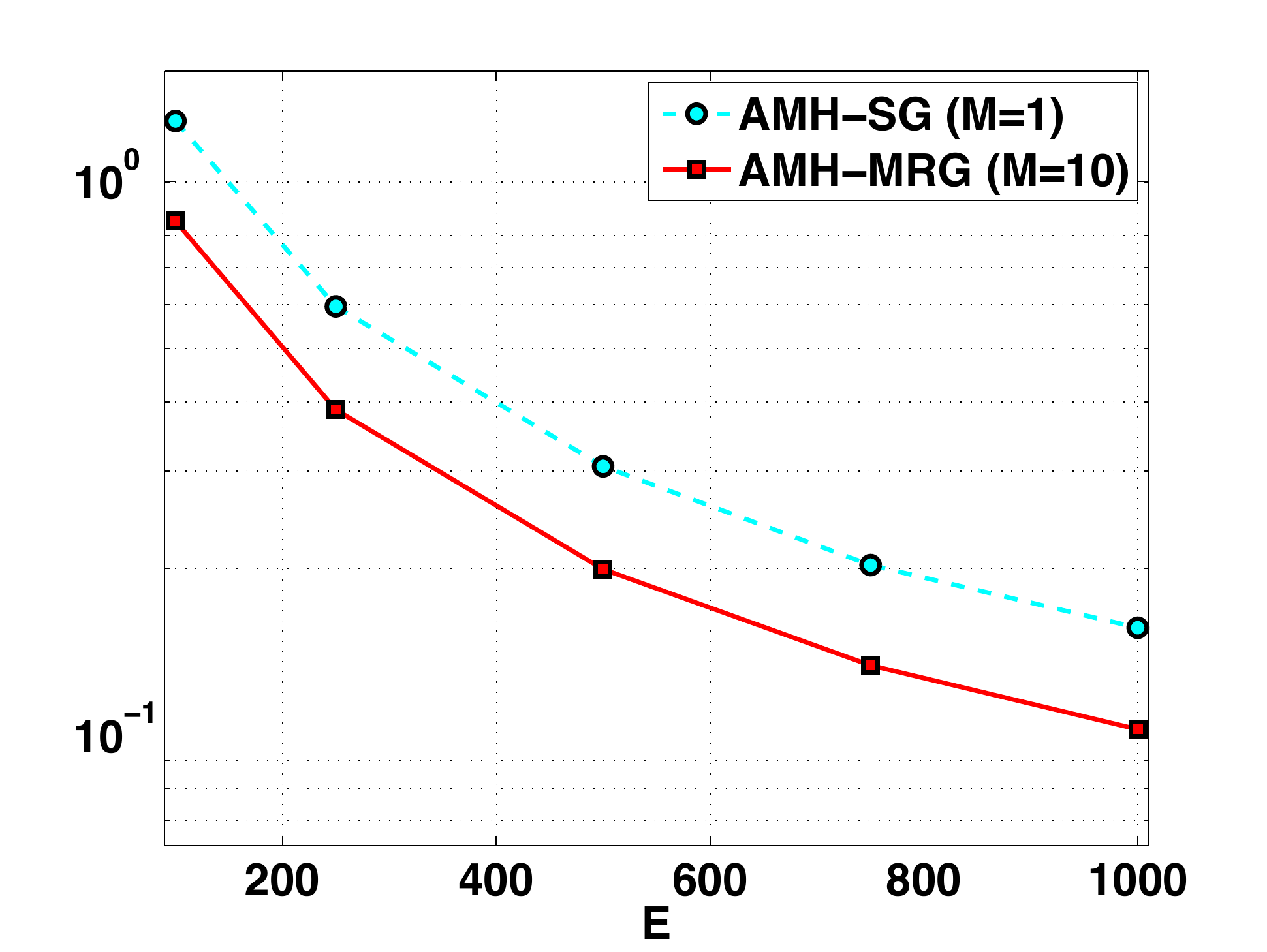}}
  \subfigure[Normalized spent time as function of $E=MT$.]{\includegraphics[width=0.5\textwidth]{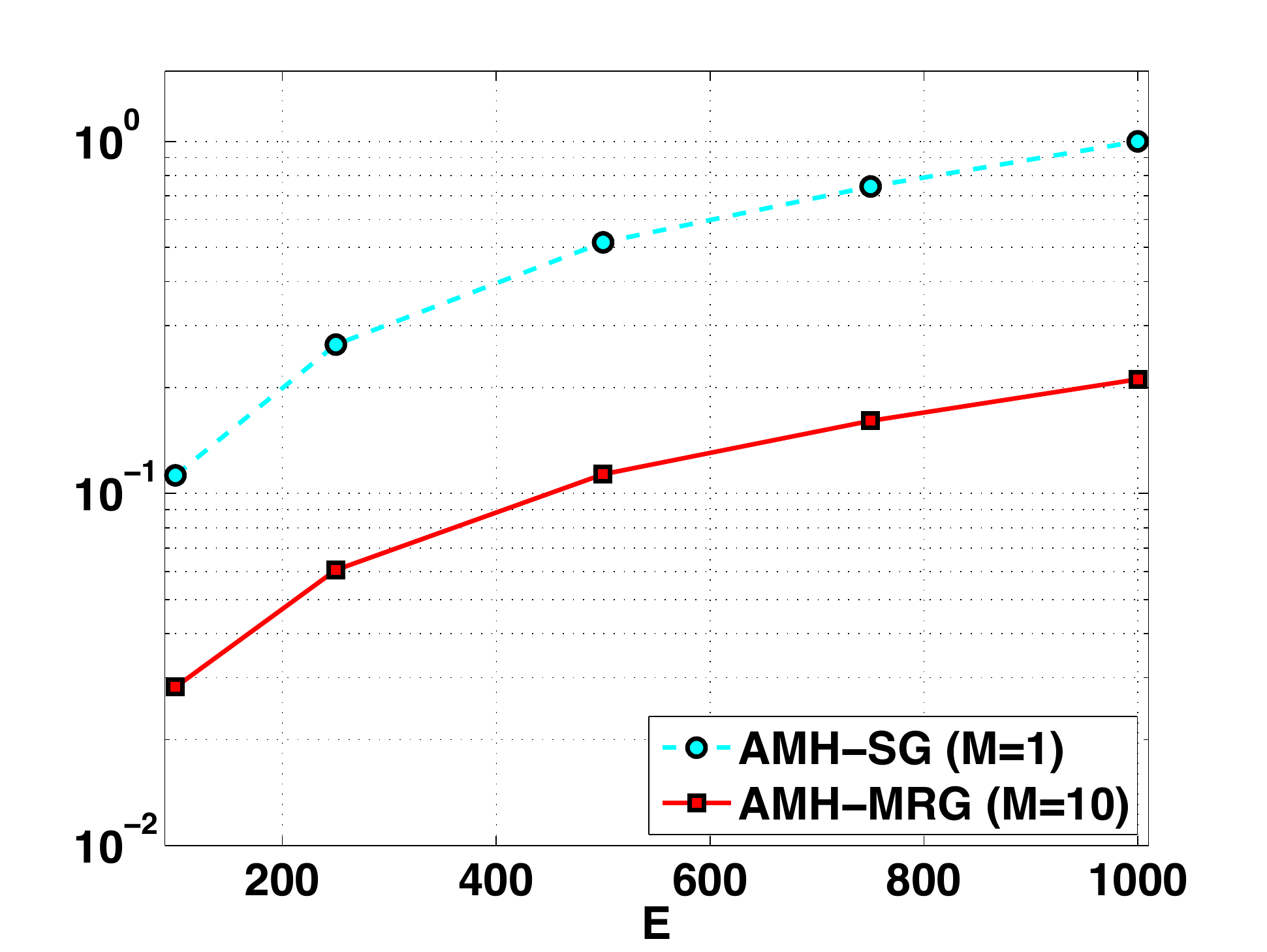}}
%  \subfigure[MSE as function of $M$ with $MT=10^4$ fixed.]{\includegraphics[width=0.5\textwidth]{Fig_MHG_MTfixed.pdf}}
}
\caption{{{\bf Exp. Section \ref{ExFirstAn}-}  {\bf (a)} MSE (log-scale) as function of the number of target evaluations per full-conditional $E=MT$, for an Adaptive MH-within-Gibbs schemes with $M=1$ and $M=10$. Clearly, for the technique with $M=10$, we use a shorter chain (smaller $T$) than for the technique with $M=1$, in order to ensure the same number of evaluations $E$. {\bf (b)} Normalized spent time (log-scale) as function of $E=MT$, normalized with respect to the time spent by AMH-SG, with $M=1$ and $T=1000$.}}
\label{FigSIMU0_2}
\end{center}
\end{figure}
}
%%%%%%%%%%%%%%%%%%%%%%%%%%%%%%%%
\subsection{Experiment 2: A second analysis of the efficiency}
%\subsection{Toy Example}
\label{ExSecAn}
%%%%%%%%%%%%%%%%%%%%%%%%%%%%%%%%
 
We test the new MRG scheme in a simple numerical simulation involving a bi-modal, bi-dimensional target pdf: %, easily reproducible by any practitioner. % in order to show that the application of the novel MRG technique is convenient even in this simple example. 
$$
{\bar \pi}(x_1,x_2)\propto \exp\left(-\frac{(x_1^2-\mu_1)^2}{2\delta_1^2}-\frac{(x_2-\mu_2)^2}{2\delta_2^2}\right),
$$
with $\mu_1=4$, $\mu_2=1$, $\delta_1=\sqrt{\frac{5}{2}}$ and $\delta_2=1$. Figure~\ref{FigTodo} shows the contour plot of ${\bar \pi}(x_1,x_2)$ and Figures \ref{FigTodo}(a)-(b) depicts some generated samples by MH-within-SG and MH-within-MRG, respectively. Figure \ref{FigTodo}(c) the corresponding histogram obtained by the MRG samples. Our goal is to approximate via Monte Carlo the expected value, ${\mathbb E}[{\bf X}]$ where ${\bf X}=[X_1,X_2] \sim {\bar \pi}(x_1,x_2) $. We test different Gibbs techniques: the MH~\cite{Robert04} and IA$^2$RMS~\cite{MartinoA2RMS} algorithms within Standard Gibbs (SG) and within MRG sampling schemes. For the MH method, we use a Gaussian random walk proposal, 
$$
q(x_{d,m}^{(t)}|x_{d,m-1}^{(t)}) \propto \exp\left(-\frac{(x_{d,m}^{(t)}-x_{d,m-1}^{(t)})^2}{2\sigma^2}\right),
$$ 
for $d \in \{1,2\}$, $1 \le m \le M$ and $1 \le t \le T$. We test different values of the $\sigma$ parameter. 
For IA$^2$RMS, we start with the set of support points  $\mathcal{S}_0=\{ -10,-6,-2,2,6,10\}$, see~\cite{MartinoA2RMS} for further details. We averaged the Mean Square Error (MSE) over $10^5$ independent runs for each Gibbs scheme. %At each run and for each element of the expected value and the variances, %(i.e., for $4$ different values), we compute the error in estimation of the true values and compute the Mean Square Error (MSE) averaged. % over the $10^5$ independent runs. 
%Then, we obtained an averaged MSE over the different four estimated values. 
 
\begin{figure}[h!]
\begin{center}
\centerline{
\subfigure[MSE as function of $\sigma$.]{ \includegraphics[width=0.5\textwidth]{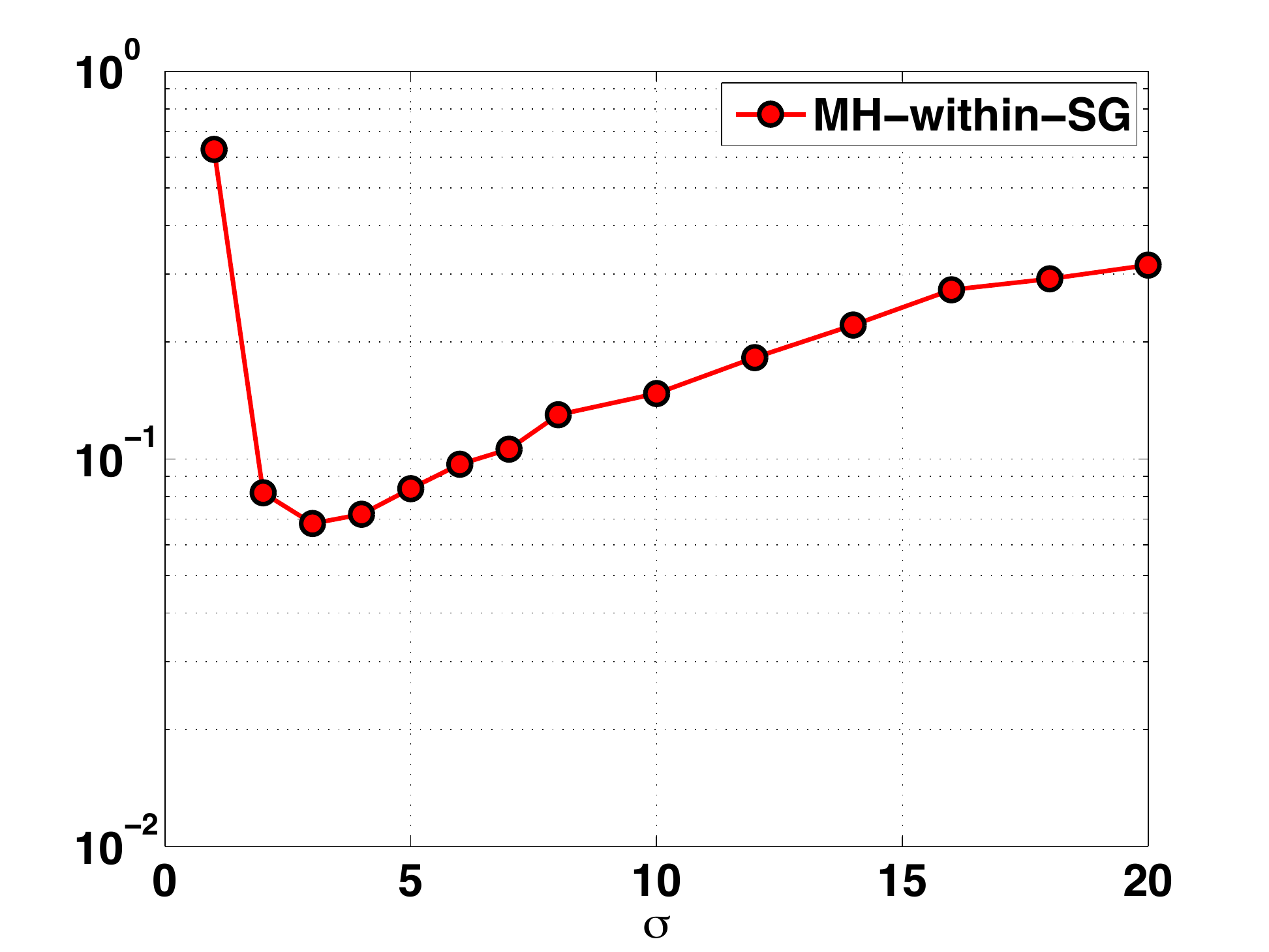}}
\subfigure[MSE as function of $T$.]{\includegraphics[width=0.5\textwidth]{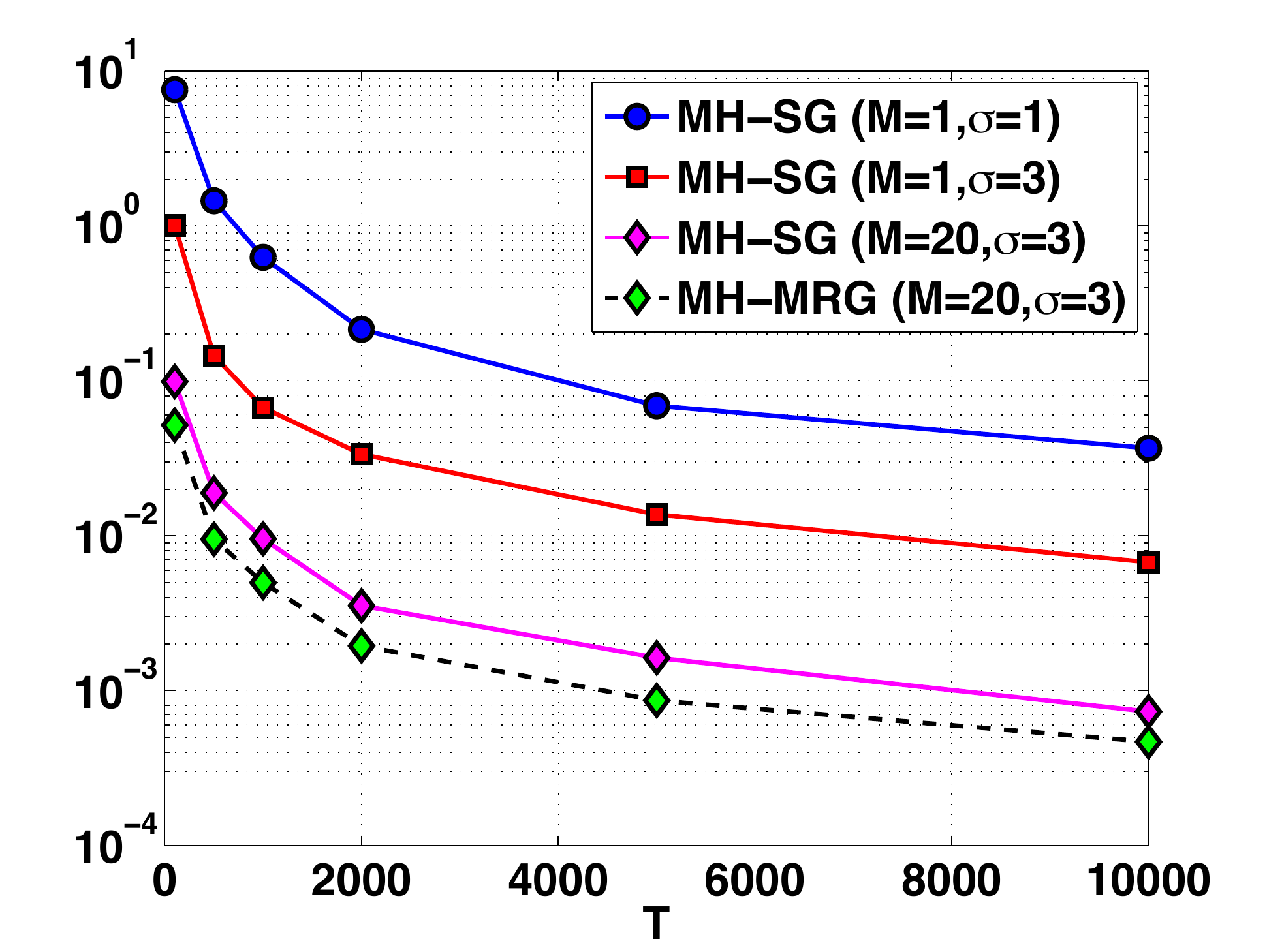}}
}
\caption{{\bf Exp. Section \ref{ExSecAn}-} {\bf (a)} MSE (log-scale) as function of $\sigma$  for MH-within-SG ($M=1$ and $T=1000$). {\bf (b)} MSE (log-scale) as function of $T$  for MH-within-SG and  MH-within-MRG schemes. We have tested $M\in\{1,20\}$ and $\sigma=\{1,3\}$ (we recall that $\sigma=3$ is the optimal scale parameter for MH; see Figure {\bf (a)}).}
\label{FigSIMU1}
\end{center}
\end{figure}

\begin{figure}[h!]
\begin{center}
\centerline{
 \subfigure[MSE as function of $M$ ($T=1000$).]{\includegraphics[width=0.5\textwidth]{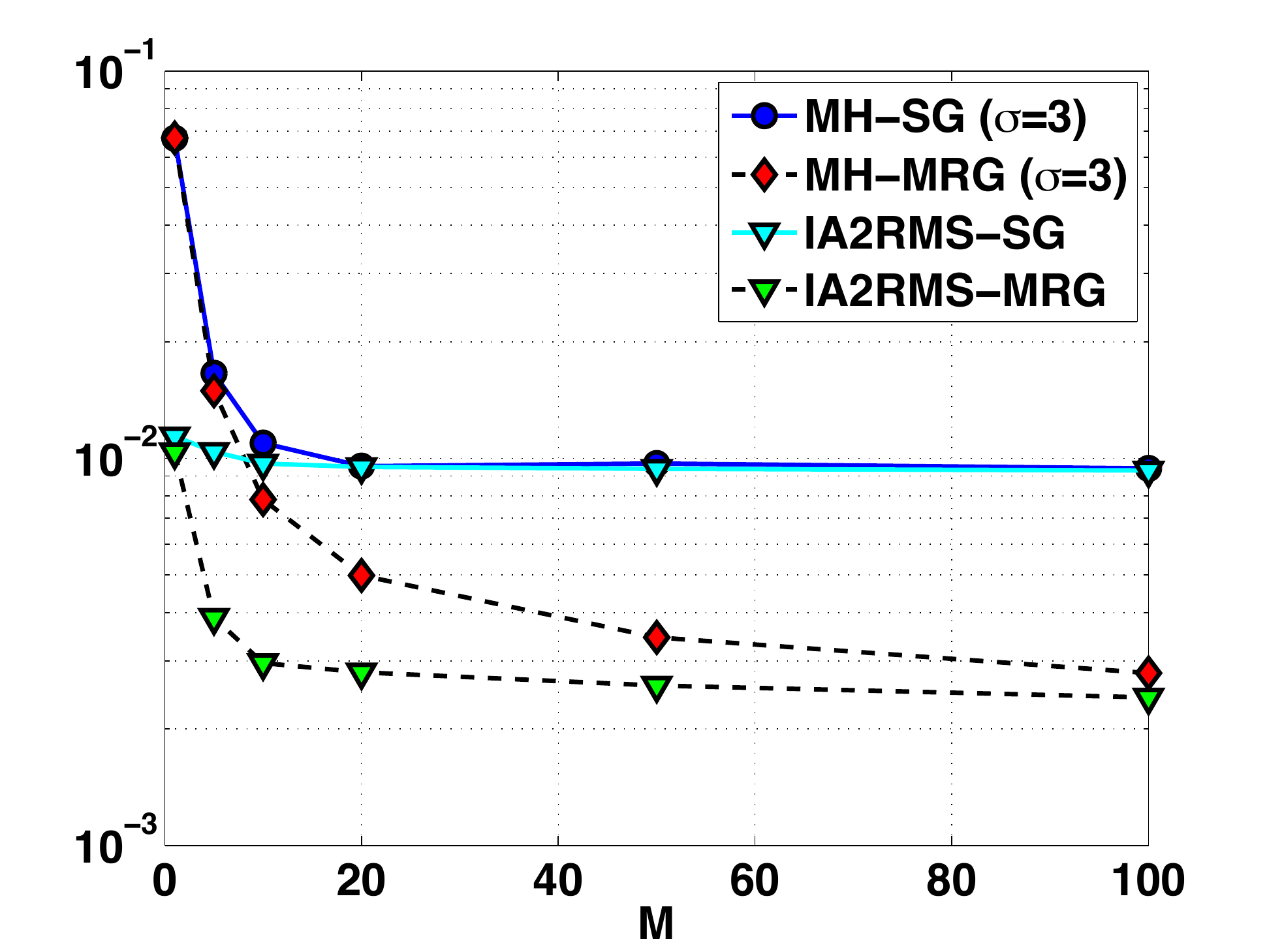}}
  \subfigure[MSE as function of spent time (sec.).]{\includegraphics[width=0.5\textwidth]{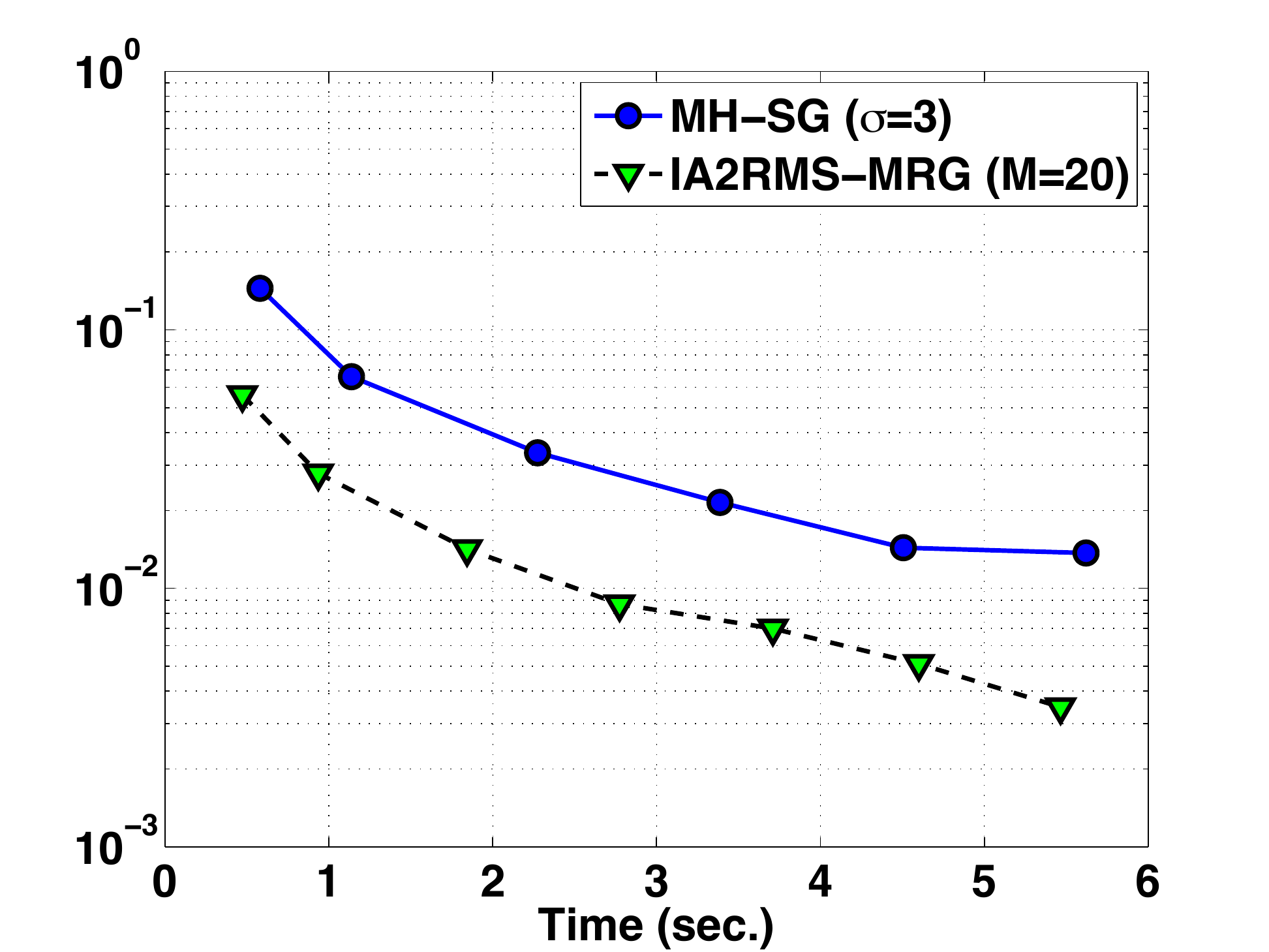}}
%  \subfigure[MSE as function of $M$ with $MT=10^4$ fixed.]{\includegraphics[width=0.5\textwidth]{Fig_MHG_MTfixed.pdf}}
}
\caption{{\bf Exp. Section \ref{ExSecAn}-} {\bf (a)} MSE (log-scale) as function of $M$ for different MCMC-within-Gibbs schemes (we fix $T=1000$).  {\bf (b)} MSE (log-scale) as function of the spent computational time (seconds). For IA$^2$RMS-within-MRG, we fix $M=20$ and vary $T$, computing the MSE in estimation and the corresponding spent computational time. For MH-within-SG, we set $\sigma=3$, $M=1$, and vary $T'$ (longer than $T$) and again we compute  MSE  and the spent time.
}
\label{FigSIMU2}
\end{center}
\end{figure}

Figure~\ref{FigSIMU1}(a) shows the MSE (in log-scale) of the MH-within-SG scheme as function of the standard deviation $\sigma$ of the proposal pdf (we set $M=1$ and $T=1000$, in this case). The performance of the Gibbs samplers depends strongly on the choice of $\sigma$ of the {\it internal} MH method. The optimal value is approximately $\sigma^*\approx 3$. The use of an adaptive proposal pdf is a possible solution, as shown in Figure~\ref{FigSIMU2}(a). Figure~\ref{FigSIMU1}(b) depicts the MSE  (in log-scale) as function of $T$ with $M=1$ and $M=20$ (for MH-within-SG we also show the case $\sigma=1$ and $\sigma=3$). Again we observe the importance of using the optimal value $\sigma^*\approx 3$ and, as a consequence, using an adaptive proposal pdf is recommended, see e.g.~\cite{Haario01}. Moreover, the use $M=20$ improves the results even without employing all the points in the estimators (i.e., in a SG scheme) since, as $M$ increases, we improve the convergence of the internal chain. Moreover, the MH-within-MRG technique provides the smallest MSE values. 

We can thus assert that recycling the internal samples provides more efficient estimators, as confirmed by Figure~\ref{FigSIMU2}(a) (represented again in log-scale). Here we fix $T=1000$ and vary $M$. As $M$ increases, the MSE becomes smaller when the MRG technique is employed. When a (SG) sampler is used, the curves show an horizontal asymptote since the internal chains converge after a certain value $M\geq M^*$, and there is not a great benefit from increasing $M$ (recall that in SG we do not recycle the internal samples). Within an MRG scheme, the increase of $M$ yields lower MSE since now we recycle the internal samples. Clearly, the benefit of using MRG w.r.t. SG increases as $M$ grows. 
Figure~\ref{FigSIMU2}(a) also shows the advantage of using an adaptive MCMC scheme (in this case IA$^2$RMS~\cite{MartinoA2RMS}). The advantage is clearer when the MH and IA$^2$RMS schemes are used within MRG. More specifically note that, as the MH method employed the optimal scale $\sigma^*\approx 3$, Figure~\ref{FigSIMU2}(a) shows the importance of a non-parametric construction of the proposal pdf employed in IA$^2$RMS. Actually, such construction allows adaptation of the entire shape of the proposal, which becomes closer and closer to the target. 
The performance of IA$^2$RMS and MH within Gibbs becomes more similar as $M$ increases. This is due to the fact that, in this case, with a high enough value of $M$, the MH chain is able to exceed its burn-in period and eventually converges. Finally, note that the adaptation speeds up the convergence of the chain generated by IA$^2$RMS. The advantage of using the adaptation is more evident for intermediate values of $M$, e.g., $10<M<30$, where the difference with the use of a standard MH is higher. As $M$ increases and the chain generated by MH converges, the difference between IA$^2$RMS and MH is reduced.
 
%{\bf Comparison with a longer Gibbs chain.} 
In Figure ~\ref{FigSIMU2}(b), we compare the performance of IA$^2$RMS-within-MRG scheme, setting $M=20$ and varying $T$, with MH-within-a standard Gibbs scheme (i.e., $M=1$) with a longer chain, i.e., a higher value of $T'>T$. In order to provide a comparison as fair as possible, we use the optimal scale parameter $\sigma^*\approx 3$ for the MH method. For each value of $T$ and $T'$, the MSE and computational time (in seconds) is given.\footnote{The computational times are obtained in a Mac processor 2.8 GHz Intel Core i5.} We can observe that, for a fixed time,  % (shown in log-scale for a proper visualization) 
IA$^2$RMS-within-MRG outperforms in MSE the standard MH-within-Gibbs scheme with a longer chain. %Namely,  IA$^2$RMS-within-MRG obtains the same MSE value spending less computational time (i.e., it is faster). 
These observations confirm the computational and estimation advantages of the proposed MRG approach.

{
%%%%%%%%%%%%%%%%%%%%%%%%%%%%%%
\subsection{Experiment 3: A third analysis of the efficiency}
\label{ExThirdAn}
%%%%%%%%%%%%%%%%%%%%%%%%%%%%%%
In this section, we consider a bi-dimensional target density which presents a strong nonlinear dependence between the variable $x_1$ and $x_2$, i.e.,
\begin{align}
\label{FullEx3}
	\bar{\pi}(x_1,x_2) & \propto \exp\left(-\frac{(x_1^2+Bx_2^2-A)^2}{4}\right), 
\end{align}
with $A=10$ and $B=0.1$. Figure \ref{FigSIMUdonut_2}(b) depicts the contour-plot of $ \bar{\pi}(x_1,x_2)$. Note that the difference scale in the first and second axis.
 We use different Monte Carlo techniques in order to approximate the expected values (groundtruth $\mu_1=0$ and $\mu_2=0$) and the standard deviations of the marginal pdfs  (groundtruth $\delta_1\approx \sqrt{5}$  and $\delta_2\approx \sqrt{51}$). We compute the average MSE in the estimation of these $4$ values (also averaged over $2000$ independent runs).

We compare two schemes, MH-within-SG and MH-within-MRG, considering again a random walk proposal pdf $q(x_{d,m}^{(t)}|x_{d,m-1}^{(t)}) \propto \exp\left(-\frac{(x_{d,m}^{(t)}-x_{d,m-1}^{(t)})^2}{2\sigma^2}\right)$, for $d \in \{1,2\}$, $1 \le m \le M$ and $1 \le t \le T$. First of all, we set $T=200$, $\sigma=10$ and vary $M$, as shown in Figure \ref{FigSIMUdonut_1}(a). Then, in Figure \ref{FigSIMUdonut_1}(b), we set $M=100$, $\sigma=10$ and vary $T$. In Figure \ref{FigSIMUdonut_2}(a), we keep fixed $M=100$, $T=200$ and change $\sigma$. The MRG schemes are depicted with solid lines whereas all the SG methods are shown with dashed lines. We can observe that the MH-within-MRG scheme always provides the best results.  Figure \ref{FigSIMUdonut_2}(a) shows again that performance of the Gibbs schemes depends on the scale parameter of the proposal. In all cases, MRG provides smaller MSE values and the benefit w.r.t. SG is more evident when a good choice of $\sigma$ is employed.

  \begin{figure}[h!]
\begin{center}
\centerline{
 \subfigure[MSE as function of $M$ ($T=200$).]{\includegraphics[width=0.5\textwidth]{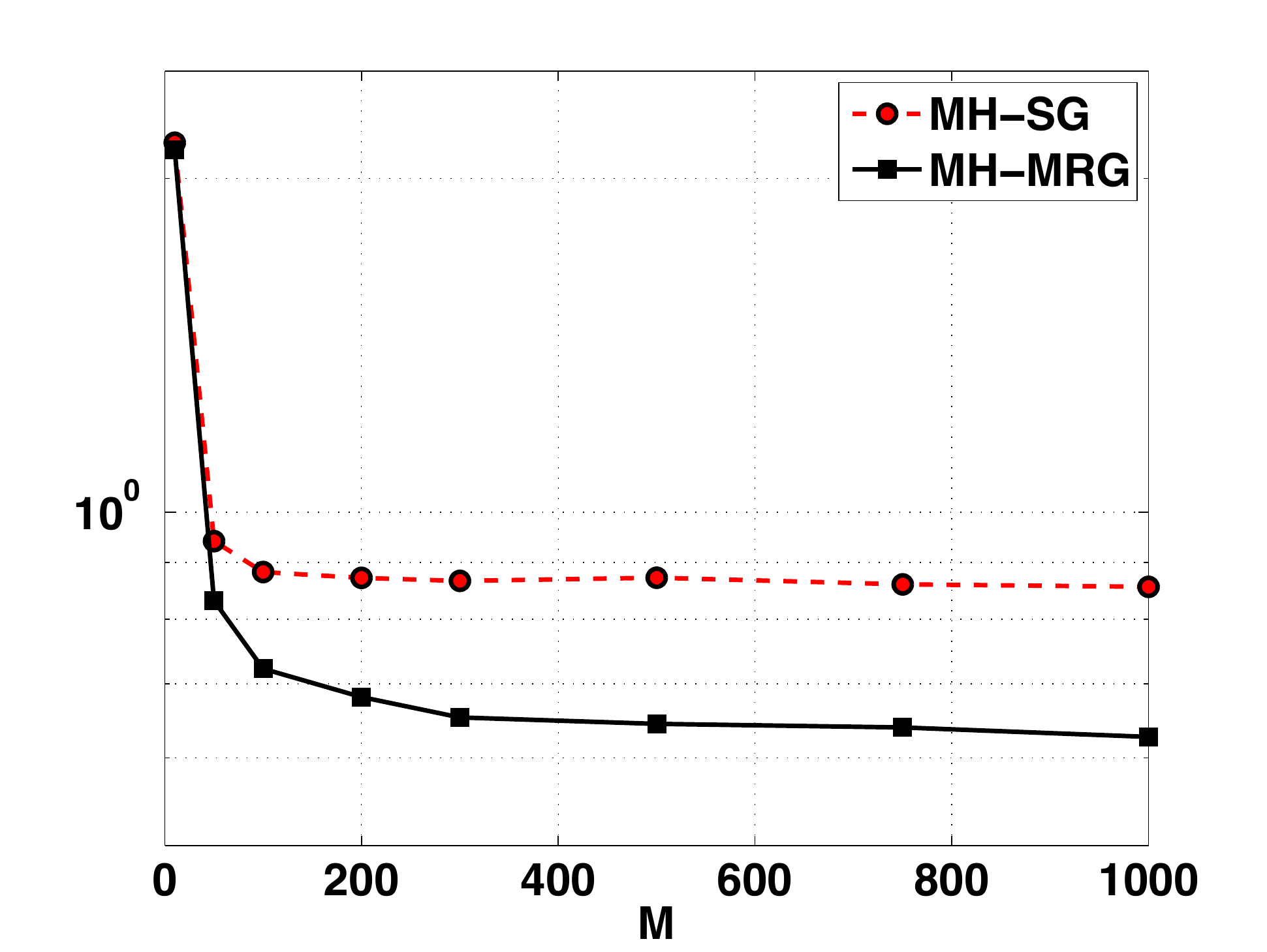}}
  \subfigure[MSE as function of $T$ ($M=100$).]{\includegraphics[width=0.5\textwidth]{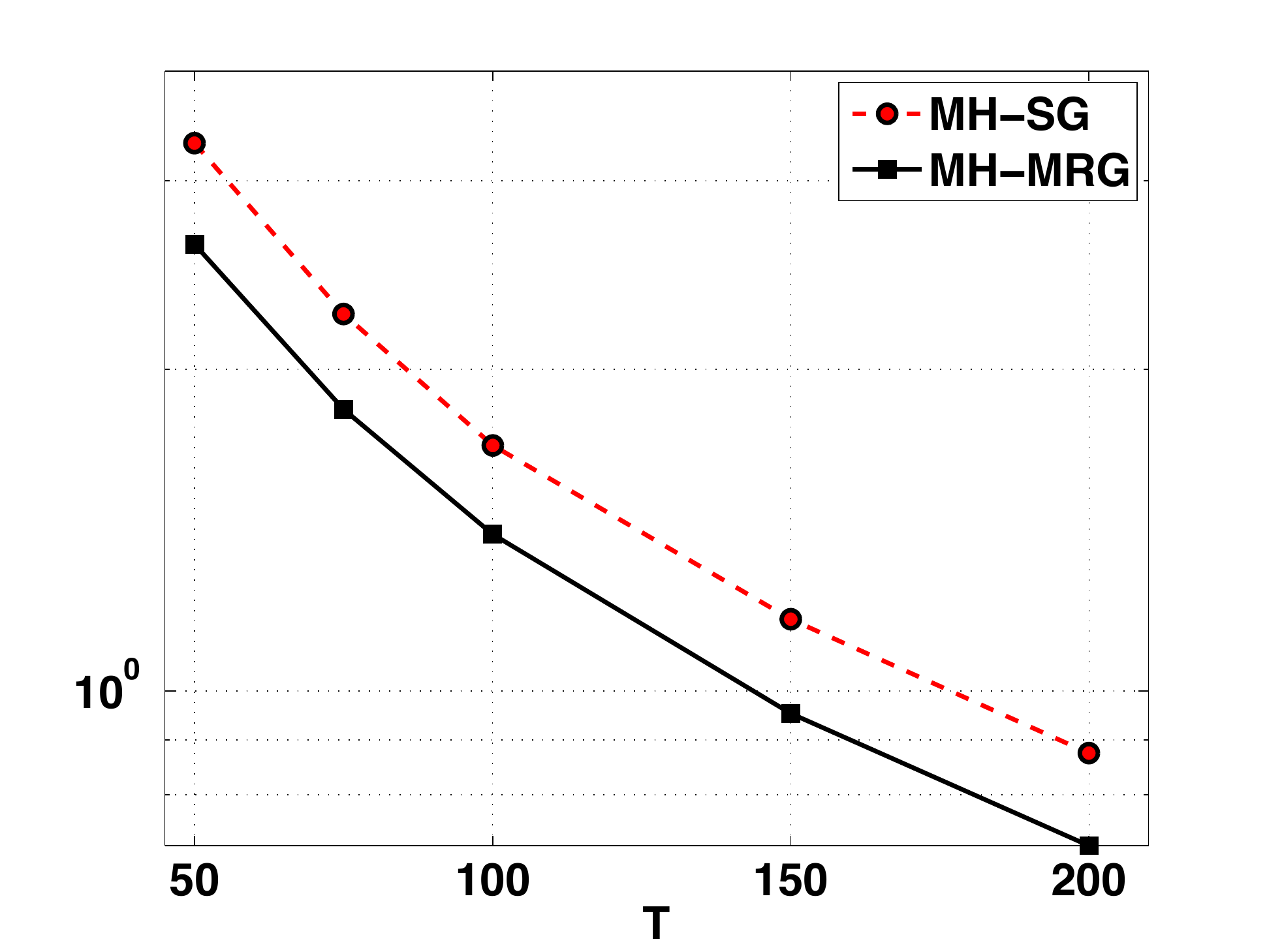}}
}
\caption{{{\bf Exp. Section \ref{ExThirdAn}-}  {\bf (a)} MSE (log-scale) as function of $M$ for MH-within-SG and MH-within-MRG  schemes with $\sigma=10$ (we fix $T=200$). {\bf (b)} MSE (log-scale) as function of $T$ keeping fixed $M=100$ and $\sigma=10$. Note that in both Figures {\bf (a)} and {\bf (b)},the MRG schemes are depicted with solid lines whereas all the SG schemes with dashed lines.}}
\label{FigSIMUdonut_1}
\end{center}
\end{figure}

  \begin{figure}[h!]
\begin{center}
\centerline{
 \subfigure[MSE as function of $\sigma$ ($M=100$, $T=200$).]{\includegraphics[width=0.5\textwidth]{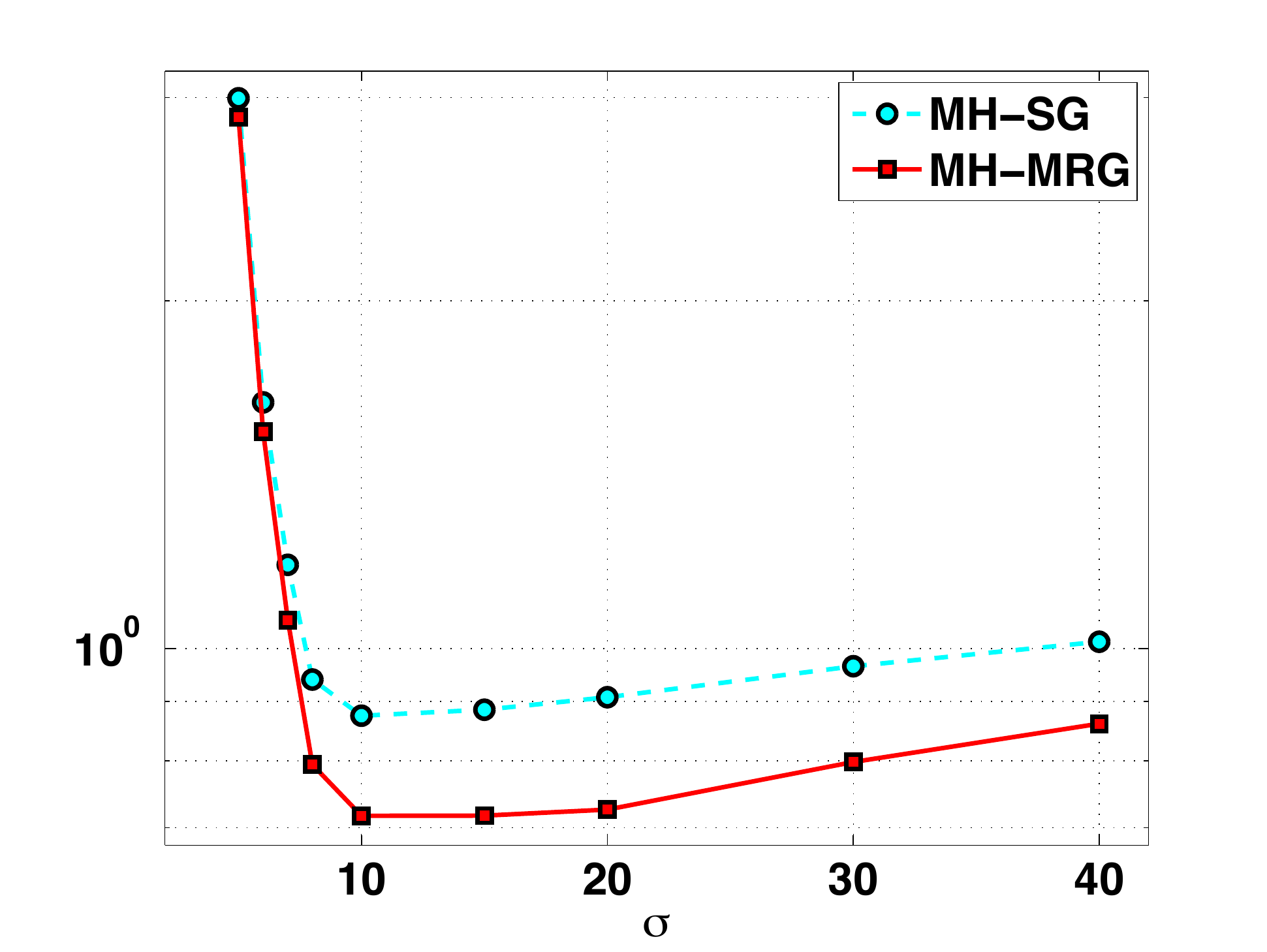}}
  \subfigure[Counterplot of the donut target pdf $\bar{\pi}$.]{\includegraphics[width=0.5\textwidth]{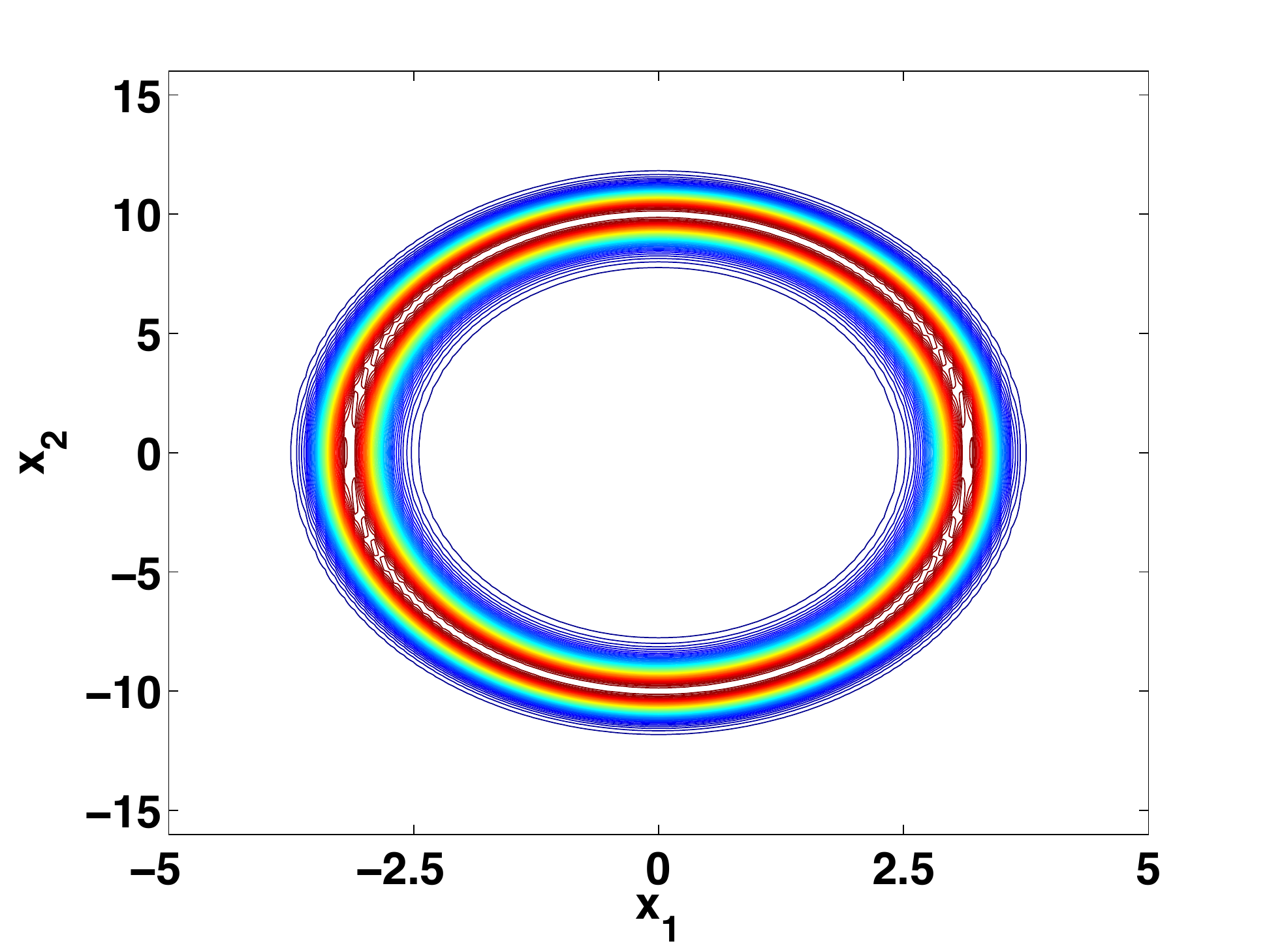}}
}
\caption{{{\bf Exp. Section \ref{ExThirdAn}-}  {\bf (a)} MSE (log-scale) as function of different values of $\sigma$ (keeping fixed $M=100$ and $T=200$). {\bf (b)} Contour-plot of the donut target pdf $\bar{\pi}$.}}
\label{FigSIMUdonut_2}
\end{center}
\end{figure}
}

%%%%%%%%%%%%%%%%%%%%%%%%%%%%%%%%%%%%%%%%%%
\subsection{Experiment 4: Learning Hyperparameters in Gaussian Processes}
%%%%%%%%%%%%%%%%%%%%%%%%%%%%%%%%%%%%%%%%%%
%\subsection{Learning Hyperparameters in Gaussian Processes Regression}
\label{GPexample}

Gaussian processes (GPs) are Bayesian state-of-the-art tools for function approximation and regression~\cite{rasmussen2006gaussian}. As for any kernel method, selecting the covariance function and learning its hyperparameters is the key to attain significant performance. We here evaluate the proposed approach for the estimation of hyperparameters of the Automatic Relevance Determination (ARD) covariance~\cite[Chapter 6]{Bishop}. 
Notationally, let us assume observed data pairs $\{y_j,{\bf z}_j\}_{j=1}^{P}$, with $y_j\in \mathbb{R}$ and
$$
{\bf z}_j=[z_{j,1},z_{j,2},\ldots,z_{j,L}]^{\top}\in \mathbb{R}^{L},
$$
 where $L$ is the dimension of the input features. We also denote the corresponding $P\times 1$ output vector as ${\bf  y}=[y_1,\ldots,y_P]^{\top}$ and the $L\times P$ input matrix  ${\bf  Z}=[{\bf z}_1,\ldots,{\bf z}_P]$. We address the regression problem of inferring the unknown function $f$ which links the variable $y$ and ${\bf z}$. Thus, the assumed model is
 \begin{equation}
 \label{ModelTrue}
y=f({\bf z})+e,
\end{equation} 
where $e\sim N(e;0,\sigma^2)$, and that $f({\bf z})$ is a realization of a Gaussian Process (GP)~\cite{rasmussen2006gaussian}. Hence $f({\bf z}) \sim \mathcal{GP}(\mu({\bf z}),\kappa({\bf z},{\bf r}))$ where $\mu({\bf z})=0$, ${\bf z},{\bf r} \in \mathbb{R}^{L}$, and we consider the ARD kernel function 
\begin{equation}
\label{EqKernel}
\kappa({\bf z},{\bf r})=\exp\left(-\sum_{\ell=1}^{L}\frac{(z_\ell-r_\ell)^2}{2\delta_\ell^2}\right), \mbox{ } \mbox{ with } \mbox{  } \delta_\ell> 0, \quad \ell=1,\ldots,L.
\end{equation}
Note that we have a different hyper-parameter $\delta_\ell$ for each input component $z_\ell$, hence we also define ${\bm \delta}=\delta_{1:L}=[\delta_1,\ldots,\delta_L]$. %This type of kernel function is often used to perform an {\it automatic relevance determination} (ARD) of the input components with respect the output variable \cite[Chapter 6]{Bishop}. Namely, 
Using ARD allows us to infer the relative importance of different components of inputs: a small value of $\delta_{\ell}$ means that a variation of the $\ell$-component $z_\ell$ impacts the output more, while a high value of $\delta_{\ell}$ shows virtually independence between the $\ell$-component and the output.

 Given these assumptions, the vector ${\bf  f}=[f({\bf z}_1),\ldots, f({\bf z}_P)]^\top$ is distributed as 
\begin{equation}
\label{Eq_f}
p({\bf  f}|{\bf  Z},{\bm \delta}, \kappa)=\mathcal{N}({\bf  f};{\bf  0},{\bf  K}),
\end{equation} 
 where ${\bf  0}$ is a $P\times 1$ null vector, and ${\bf  K}_{ij}:=\kappa({\bf z}_i,{\bf z}_j)$, for all $i,j=1,\ldots,P$, is a $P\times P$ matrix. Note that, in Eq.~\eqref{Eq_f}, we have expressed explicitly the dependence on the input matrix ${\bf Z}$, on the  vector  ${\bm \delta}$ and on the choice of the kernel family $\kappa$.
 Therefore, the vector containing all the hyper-parameters of the model  is
 \begin{eqnarray*}
 {\bm \theta}&=&[\theta_{1:L}=\delta_{1:L},\theta_{L+1}=\sigma], \\
  {\bm \theta}&=&[{\bm \delta}, \sigma] \in \mathbb{R}^{L+1},
\end{eqnarray*}
i.e., all the parameters of the kernel function in Eq.~\eqref{EqKernel} and standard deviation $\sigma$ of the observation noise.  
Considering the filtering scenario and the tuning of the parameters (i.e., inferring the vectors ${\bf  f}$ and ${\bm \theta}$), the full Bayesian solution addresses the study of the full posterior pdf involving  ${\bf  f}$ and ${\bm \theta}$,
\begin{equation}
\label{CompletePost}
p({\bf  f},{\bm \theta}|{\bf  y}, {\bf  Z}, \kappa)=\frac{p({\bf  y}|{\bf  f},{\bf Z},{\bm \theta}, \kappa)p({\bf  f}|{\bf  z},{\bm \theta},\kappa) p({\bm \theta})}{p({\bf  y}|{\bf  Z},\kappa)},
\end{equation}
where $p({\bf  y}|{\bf  f},{\bf  Z},{\bm \theta}, \kappa)=\mathcal{N}({\bf  y};{\bf  0},\sigma^2 {\bf  I})$ given the observation model in Eq.~\eqref{ModelTrue},  $p({\bf  f}|{\bf  z},{\bm \theta},\kappa)$ is given in Eq.~\eqref{Eq_f}, and $p({\bm \theta})$ is the prior over the  hyper-parameters.  We assume $p({\bm \theta})=\prod_{\ell=1}^{L+1}\frac{1}{\theta_\ell^{\beta}}\mathbb{I}_{\theta_\ell}$ where $\beta=1.3$ and $\mathbb{I}_{v}=1$  if $v>0$, and $\mathbb{I}_{v}=0$  if $ v\leq 0$. Note that the posterior in Eq.~\eqref{CompletePost} is analytically intractable but, given a fixed vector ${\bm \theta}'$, the marginal posterior of $p({\bf f}|{\bf  y}, {\bf  Z}, {\bm \theta}',\kappa)=\mathcal{N}({\bf f}; {\bm \mu}_p,{\bm \Sigma}_p)$ is known in closed-form: it is Gaussian with mean $  {\bm \mu}_p={\bf  K}({\bf  K}+\sigma^2 {\bf  I})^{-1} {\bf  y}$ and covariance matrix ${\bm \Sigma}_p={\bf  K}-{\bf  K}({\bf  K}+\sigma^2 {\bf  I})^{-1} {\bf K}$~\cite{rasmussen2006gaussian}. For the sake of simplicity, in this experiment we focus on the marginal posterior density of the hyperparameters,
$$
p({\bm \theta}|{\bf  y}, {\bf  Z}, \kappa)=\int p({\bf  f},{\bm \theta}|{\bf  y}, {\bf  Z}, \kappa) d{\bf  f}\propto p({\bf  y}|{\bm \theta}, {\bf  Z}, \kappa) p({\bm \theta}), 
$$
which can be evaluated analytically. Actually, since $p({\bf  y}|{\bm \theta}, {\bf  Z}, \kappa)=\mathcal{N}({\bf  y};{\bf  0},{\bf K}+\sigma^2 {\bf  I})$ and $p({\bm \theta}|{\bf  y}, {\bf  Z}, \kappa) \propto p({\bf  y}|{\bm \theta}, {\bf  Z}, \kappa)p({\bm \theta})$, we have 
\begin{eqnarray}
\log \left[p({\bm \theta}|{\bf  y}, {\bf  Z}, \kappa)\right]
&\propto& -\frac{1}{2} {\bf  y}^{\top} ({\bf  K}+\sigma^2 {\bf  I})^{-1} {\bf  y}-\frac{1}{2} \log\left[\mbox{det}\left[{\bf  K}+\sigma^2 {\bf  I}\right]\right]-\beta \sum_{\ell=1}^{L+1} \log\theta_\ell, 
\end{eqnarray}
with $\theta_\ell >0$, where clearly ${\bf  K}$ depends on $\theta_{1:L}=\delta_{1:L}$ and recall that $\theta_{L+1}=\sigma$~\cite{rasmussen2006gaussian}. % (we use $\beta=1.3$).  
However, the moments of this marginal posterior cannot be computed analytically. Then, in order to compute the Minimum Mean Square Error (MMSE) estimator, i.e., the expected value ${\mathbb E}[{\bm \Theta}]$ with 
${\bm \Theta} \sim p({\bm \theta}|{\bf  y}, {\bf  Z}, \kappa)$, we approximate ${\mathbb E}[{\bm \Theta}]$ via Monte Carlo quadrature. More specifically, we apply a Gibbs-type samplers to draw from $\pi({\bm \theta})\propto p({\bm \theta}|{\bf  y}, {\bf  Z}, \kappa)$. Note that dimension of the problem is $D=L+1$ since ${\bm \theta}\in \mathbb{R}^{D}$.

We generated the $P=500$ pairs of data, $\{y_j,{\bf z}_j\}_{j=1}^{P}$, drawing ${\bf z}_j\sim\mathcal{U}([0,10]^L)$ and ${\bf y}_j$ according to the model in Eq.~\eqref{ModelTrue}, considered $L\in\{1,3\}$ so that $D\in\{2,4\}$, and set $\sigma^*=\frac{1}{2}$ for both cases, $\delta^*=1$ and ${\bm \delta}^*=[1,3,1]$, respectively (recall that ${\bm \theta}^*=[{\bm \delta}^*,\sigma^*]$). Keeping fixed the generated data for each scenario, we then computed the ground-truths using an exhaustive and costly Monte Carlo approximation, in order to be able to compare the different techniques. 

\begin{figure}[h!]
\begin{center}
\centerline{
\subfigure[MSE  as function of $M$ ($T=100$).]{\includegraphics[width=0.5\textwidth]{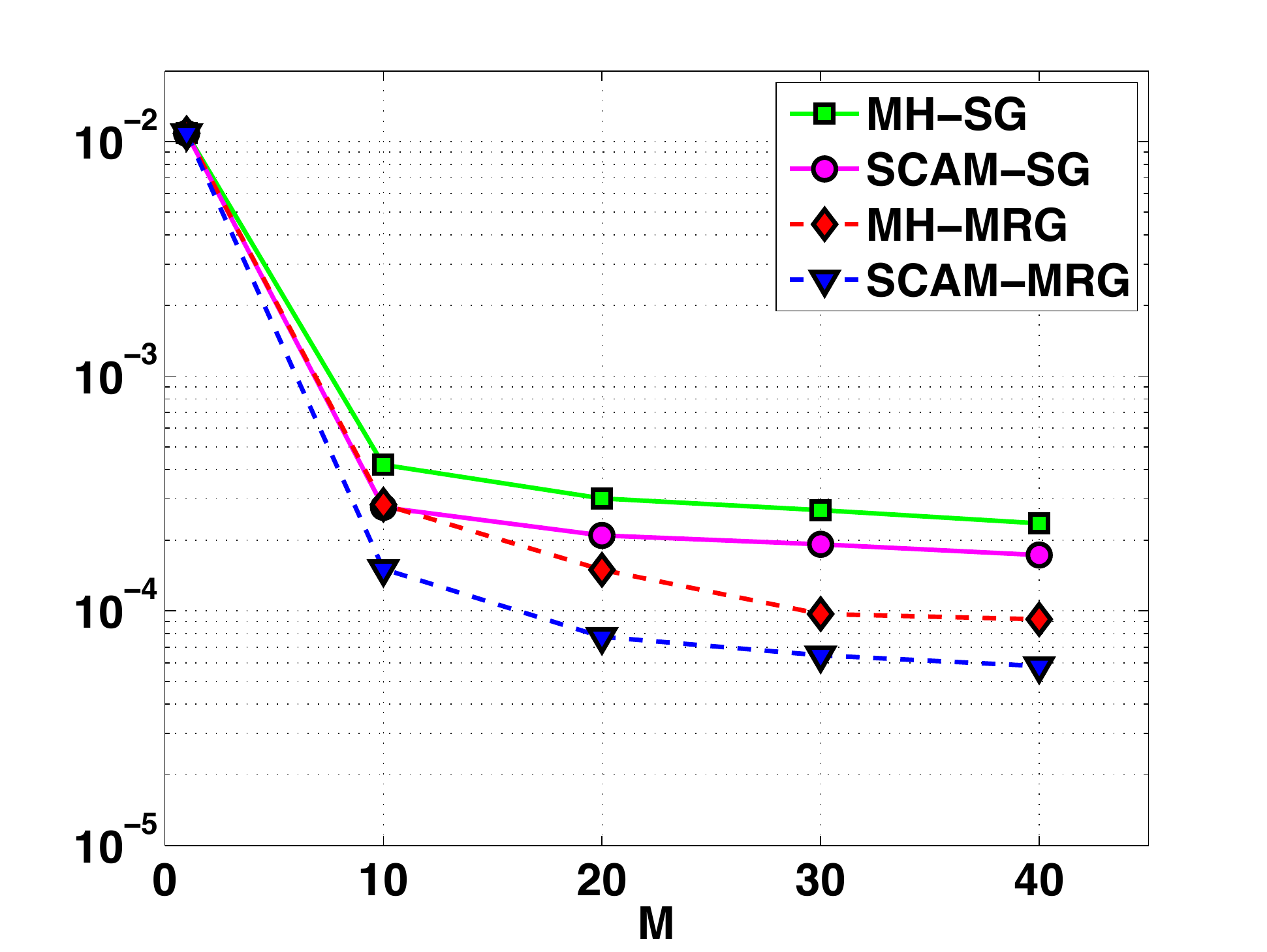}}
\subfigure[MSE  as function of $T$.]{\includegraphics[width=0.5\textwidth]{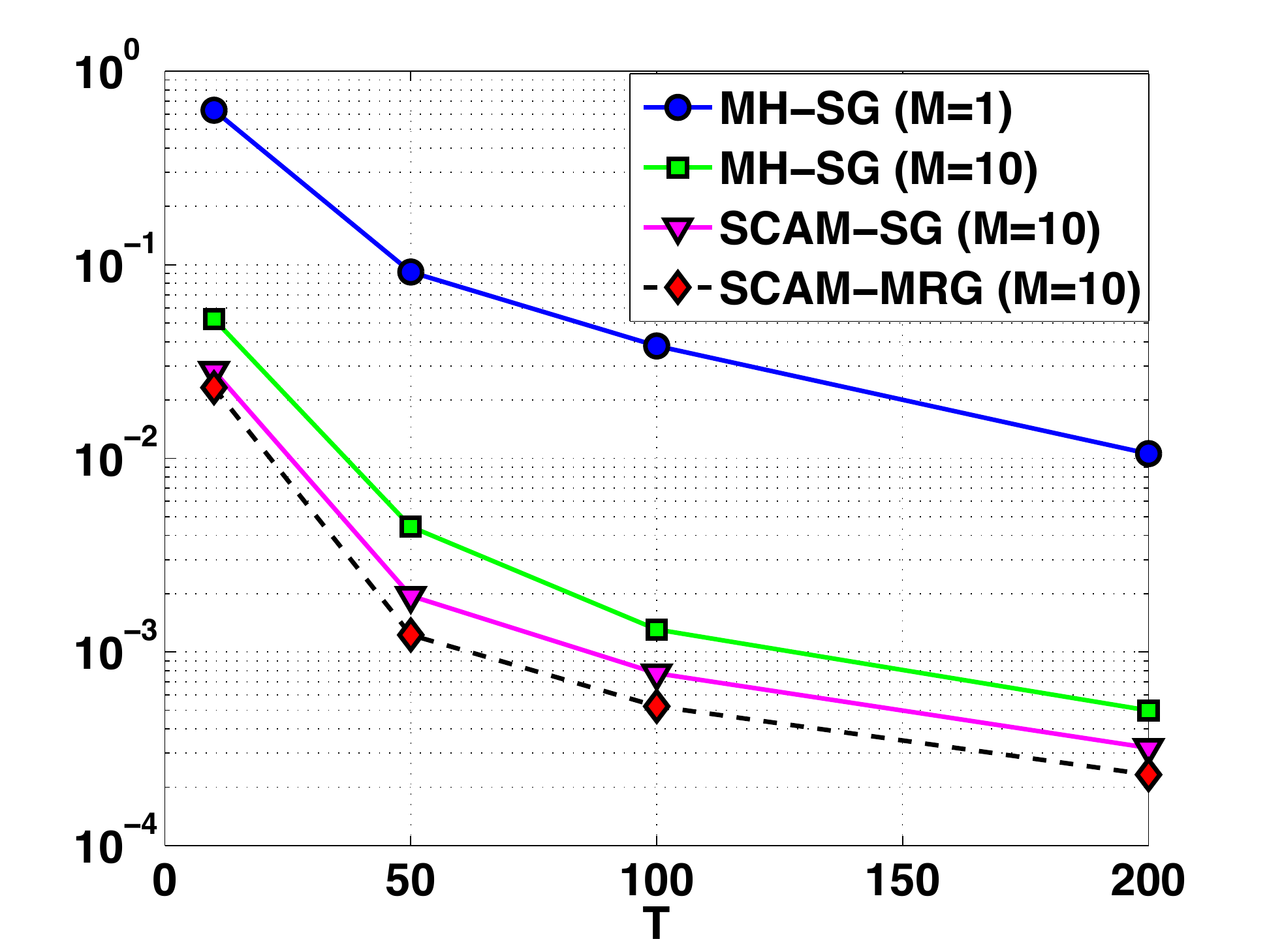}}
%  \subfigure[MSE as function of $M$ with $MT=10^4$ fixed.]{\includegraphics[width=0.5\textwidth]{Fig_MHG_MTfixed.pdf}}
}
\caption{{\bf Exp. 2-} {\bf (a)} MSE (log-scale) as function of $M$ (starting with $M=1$) for different MCMC-within-Gibbs schemes ($T=100$ and $D=2$).  {\bf (b)} MSE (log-scale) as function of $T$  for different techniques (in this case, $D=4$), with $M=1$ for the MH-within-SG method depicted with a solid line and circles, whereas $M=10$ for the remaining curves. Note that, in both figures, the MRG approaches, shown with dashed lines, always outperform the corresponding standard Gibbs (SG) schemes, shown with solid lines. 
}
\label{FigSIMU3}
\end{center}
\end{figure}

We tested the standard MH within SG and MRG {(with a random walk Gaussian proposal $q(x_{d,m}^{(t)}|x_{d,m-1}^{(t)}) \propto \exp\left(-\frac{(x_{d,m}^{(t)}-x_{d,m-1}^{(t)})^2}{2\sigma^2}\right)$, as in the previous examples, with $\sigma=2$)}, and also the Single Component Adaptive Metropolis (SCAM) algorithm~\cite{HaarioCW} within SG and MRG. SCAM is a component-wise version of the adaptive MH method~\cite{Haario01} where the covariance matrix of the proposal is automatically adapted. In SCAM, the covariance matrix of the proposal is diagonal and each element is adapted considering only the corresponding component: that is, the variances of the marginal densities of the target pdf are estimated and used as a scale parameter of the proposal pdf in the corresponding component.\footnote{More specifically, we have implemented an accelerated version of SCAM which takes  more advantage of the MRG scheme, since the variance is also adjusted online during the sampling of the considered full-conditional (for more details, see the code at \url{http://isp.uv.es/code/RG.zip}).} 
We averaged the results using $10^3$ independent runs. Figure~\ref{FigSIMU3}(a) shows the MSE curves (in log-scale) of the different schemes as function of $M\in\{1,10,20,30,40\}$, while keeping fixed $T=100$ (in this case, $D=2$). Figure~\ref{FigSIMU3}(b) depicts the MSE curves ($D=4$) as function of $T$ considering in one case $M=1$ and $M=10$ for the rest. %The MRG schemes are represented with dashed lines whereas the standard Gibbs (SG) approaches are illustrated with solid lines. 
In both figures, we notice that (1) using an $M>1$ is advantageous in any case (SG or MRG), (2) using a procedure to adapt the proposal improves the results, and (3) MRG, i.e., recycling all the generated samples, always outperforms the SG schemes. 

\begin{figure}[h!]
\begin{center}
\centerline{
 \subfigure[MSE versus the number of target evaluations $E$.]{\includegraphics[width=0.5\textwidth]{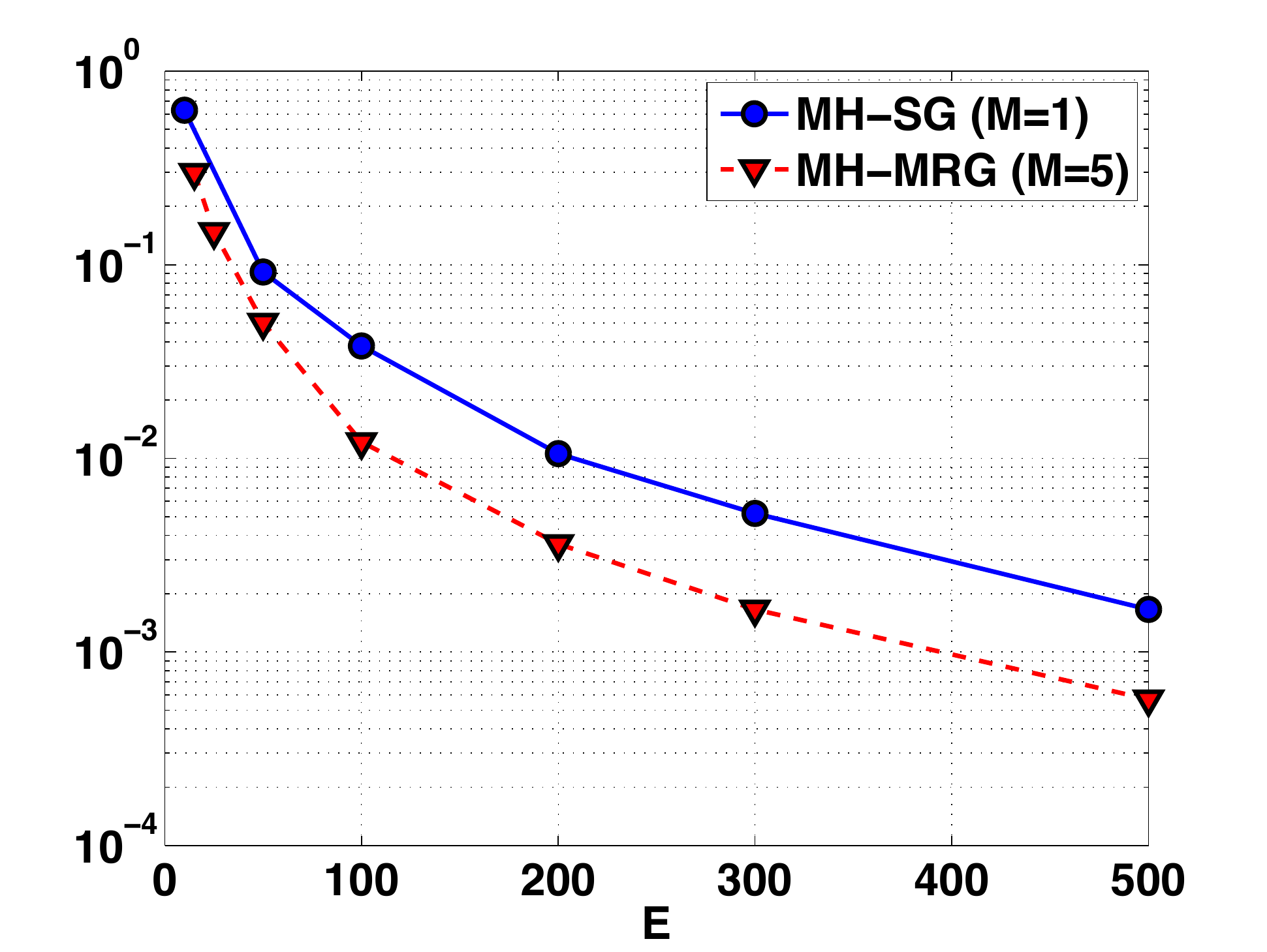}}
  \subfigure[Histograms of the generated samples.]{\includegraphics[width=0.5\textwidth]{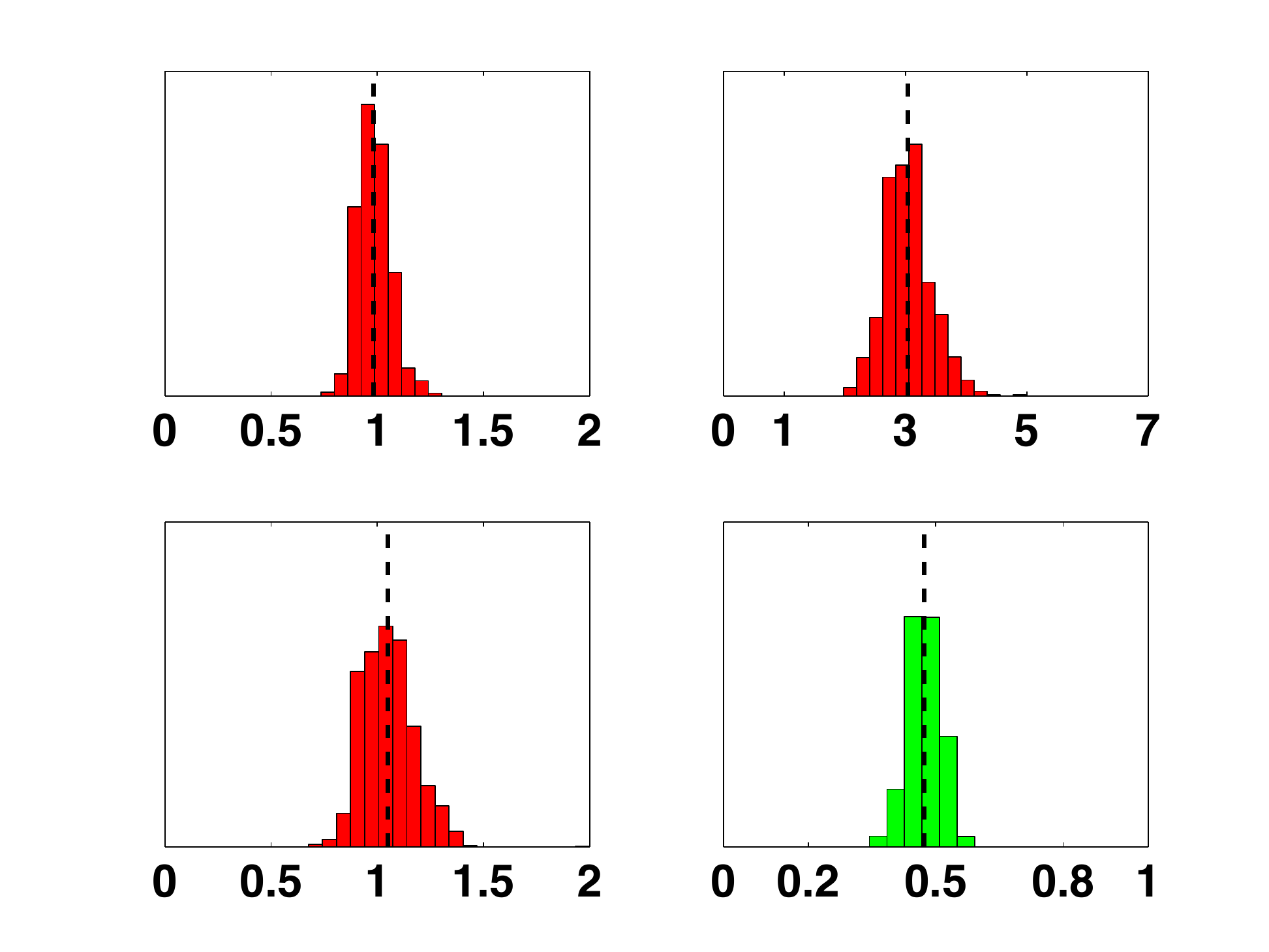}}
%  \subfigure[MSE as function of $M$ with $MT=10^4$ fixed.]{\includegraphics[width=0.5\textwidth]{Fig_MHG_MTfixed.pdf}}
}
\caption{{\bf Exp. 2-} {\bf (a)} MSE (log-scale) as function of  the total number of target evaluations $E=MT$, per full-conditional ($D=4$). Namely, for MH-within-SG we have $M=1$ and $T\in\{10,50,100, 200, 300, 500\}$, whereas for MH-within-MRG we have $M=5$ and $T\in\{3,5,10, 20, 40, 60, 100\}$.  
  {\bf (b)} Histograms of the samples drawn from the posterior $p({\bm \theta}|{\bf  y}, {\bf  Z}, \kappa)$ in a specific run, with ${\bm \theta}=[\delta_1,\delta_2,\delta_3,\sigma]$, i.e.,  $D=4$, generated using MH-within-MRG with $M=5$ and $T= 2000$.  %Recall that the data are generated according to the model with parameter values $\delta_1^*=1$, $\delta_2^*=3$, $\delta_3^*=1$ and $(\sigma^2)^*=0.5$.
}
\label{FigSIMU4}
\end{center}
\end{figure}

Figure~\ref{FigSIMU4}(a) compares the MH-within-SG with the MH-within-MRG, showing the MSE as function of the total number of target evaluations $E=MT$ per full-conditional.
We set $M=5$, $T\in\{3, 5, 10, 20, 40, 60, 100\}$ for  MH-within-MRG, whereas  we have $M=1$ and $T\in\{10, 50,$ $100, 200, 300, 500\}$ for MH-within-SG. Namely, we used a longer Gibbs chain for MH-within-SG. Note that the MH-within-MRG provides always smaller MSEs, considering the same total number of evaluations $E$ of the target density. Figure~\ref{FigSIMU4}(b) depicts the histograms of the samples ${\bm \theta}^{(t)}$  drawn from the posterior $p({\bm \theta}|{\bf  y}, {\bf  Z}, \kappa)$ in a specific run, with $D=4$, generated using MH-within-MRG with $M=5$ and $T= 2000$. The dashed line represents the mean of the samples (recall that $\delta_1^*=1$, $\delta_2^*=3$, $\delta_3^*=1$ and $(\sigma^*)^2=0.5$). Note that all the samples ${\bm \theta}^{(t)}$ can be employed for approximating the full Bayesian solution of the GP which involves the joint posterior pdf in Eq.~\eqref{CompletePost}. 

{ We recall that the results in Figure \ref{FigSIMU3}(b) and Figures \ref{FigSIMU4} corresponds to $D=4$.
We have also tested SG and MRG in higher dimensions as shown in Table \ref{tabDiffD}. In order to obtain these results (averaged over $10^3$ runs), we generate  $P=500$ pairs of data (as described above; we keep them fixed in all the runs), considering again $\sigma^*=\frac{1}{2}$ and $\delta_\ell=2$ for all $\ell=1,\ldots,L$, with $L\in\{5,7,9\}$. Hence, ${\bm \theta}^*=[{\bm \delta}^*,\sigma^*]$ has dimension $D\in\{6,8,10\}$.  We test MH-within-SG and MH-within-MRG with the random walk Gaussian proposal described above ($\sigma=2$), $M=10$ and $T=2000$. Observing Table \ref{tabDiffD}, we see that the MRG provides the best results for all the dimension $D$ of the inference problem. 
\begin{table}[!h]
\centering
{
%\small
%\footnotesize
\caption{MSE as function of the dimension $D$ (with $M=10$ and $T=2000$).  }
\vspace{0.1cm}
	\begin{tabular}{|c|c|c|c|}
    \hline
%\begin{enumerate}
%\item 
{\bf Method} &  $D=6$ &   $D=8$ &  $D=10$   \\ 
   \hline
 MH-within-SG & 0.1247   &  0.3816   &   2.0720    \\ 
 MH-within-MRG & 0.0681    & 0.1759     &  1.4681    \\ 
\hline 
\end{tabular}
}
\label{tabDiffD}
\end{table}
}

%%%%%%%%%%%%%%%%%%%%%%%%%%%%%%%%%%%%%%%%%%%%%%%%%%
\subsection{Experiment 5: Learning Dependencies in Remote Sensing Variables}
%\subsection{Learning Variable Dependencies in Remote Sensing Data}
%%%%%%%%%%%%%%%%%%%%%%%%%%%%%%%%%%%%%%%%%%%%%%%%%%
\label{Ex3}
We now consider the application of the MRG scheme to study the dependence among different geophysical variables { (considering real data).} Specifically, we consider the case of temperature estimation from thermal infrared (TIR) remotely sensed data. In this scenario, land surface temperature ($T_s$) and emissivity ($\epsilon$) are the two main geo-biophysical variables to be retrieved from TIR data, since most of the energy detected by the sensor in this spectral region is directly emitted by the land surface. The atmosphere status can be considered as a mediating variable in the relations between the satellite measured $T$ and $T_s$, and is here summarized by the integral of the water vapour $W$ through the whole atmospheric column. Both variables $T_s$ and $\epsilon$ are coupled and constitute a typical problem in remote sensing referred as to the ``temperature and emissivity separation problem''. On the one hand, models for estimating land temperature $T_s$ typically involve simple parametrizations of at-sensor brightness temperatures $T$, the mean and/or differential emissivities (${\bar \epsilon}$ and $\Delta \epsilon$), and the total atmospheric water vapour content $W$. On the other hand, a plethora of models for estimating $\epsilon$ have been devised~\cite{Jimenez08}.% Estimating surface emissivities is difficult and involves complex radiative inversion models~\cite{Sobrino08}.

Here we focus on the application of the MRG sampler to infer the statistical dependencies between the observed variables. We aim to obtain the dependence graph between the considered geophysical variables,  $X_1=T_s$, $X_2=W$, $X_3=\epsilon$ and $X_4=T$.
To assess such relations, we considered synthetic data simulating ASTER sensor conditions~\cite{Sobrino08}. A total of $6588$ data points was available. For simplicity, we focused on band 10 ($\sim 8.3\mu$m) for $T$ and $\epsilon$, and subsampled the dataset to finally work with $220$ data points. The data was subsequently standardized. 
%%%%%%%%%%%%%%%%%%
\subsubsection{Main procedure} 
%%%%%%%%%%%%%%%%%%%
We study the $12$ possible regression models of type
$$
 X_i=f_{j,i}(X_j)+E_{j,i}. \quad \quad  i,j\in\{1,2,3,4\}, \quad \mbox{ with } i\neq j, 
$$
where $E_{j,i}\sim\mathcal{N}(0,\sigma_{i,j}^2)$ and $f_{j,i}(x_j)$ is a realization of a Gaussian Process (GP)~\cite{rasmussen2006gaussian}, with zero mean and kernel function defined in Eq.~\eqref{EqKernel} (note that in this case $L=1$). For each regression problem, we trained the corresponding GP model using SCAM-within-MRG with $T=200$ and $M=10$. We analyze the empirical distributions of corresponding hyper-parameters ${\bm \theta}_{i,j}=[\delta_{i,j}, \sigma_{i,j}^2]$ obtained by Monte Carlo. We focus mainly on the distribution of $\delta_{i,j}$, i.e., the hyper-parameter of the kernel in Eq.~\eqref{EqKernel}. 
% The priors over the hyper-parameters are the same pdfs described in the previous section.
%{\color{magenta} Clearly, hyper-parameter $\delta_{i,j}$ tends to be higher and its distribution exhibits a heavier right tail if the Signal-Noise-Ratio (SNR) is low and if the dependence between $X_i$ and $ X_j$ is weak or they are independent. Roughly speaking, if the $X_i$ and $X_j$ are independent the distribution of $\delta_{i,j}$  presents an heavy right tail, with great values of central tendency measures (e.g., mean, median and mode) and also huge values of dispersion measures (e.g., variance). However, spread huge values of $\delta_{i,j}$ depends on (a) the noise power in the system, (b) the distance among the inputs and the total number of inputs, (c) the unknown mapping (deterministic or stochastic) which links $X_i$ and $X_j$, and (d) the regression order, i.e., if $X_i$ as input and $X_j$ as output, or viceversa. For these reasons, we need to construct some criterion in order to determine the significance of the obtained results. LO VEO MUY VERBOSE, HABLAMOS Y LO SIMPLIFICO.}   
Hyper-parameter $\delta_{i,j}$ will tend to be higher and its distribution will exhibit heavier right tail if the Signal-Noise-Ratio (SNR) is low and if $X_i$ and $ X_j$ are close to independence. However, the spread of $\delta_{i,j}$ also depends on the noise power in the system, the unknown mapping (deterministic or stochastic) linking $X_i$ and $X_j$, and the asymmetry of the regression functions, i.e. in general $f_{i,j}\neq f_{j,i}$. Hypothesis testing comes into play to determine the significance of the association between all pairs of variables $X_i$ and $X_j$.
%%%%%%%
\subsubsection{Hypothesis testing and surrogate data}
%%%%%
In order to find significance levels and thresholds about the existence of any possible dependence (strong or weak), we perform an hypothesis test with the null-hypothesis ``$\mathcal{H}_0$: independence between $X_i$ and $X_j$''. %{(and the alternative hypothesis $\mathcal{H}_1$ is ``there exists a deterministic link between $X_i$ and $X_j$ '')}. 
We build the sampling distribution of $\mathcal{H}_0$ by the surrogate data method in~\cite{Theiler92}. Under the null-hypothesis $\mathcal{H}_0$ of independence, some proper surrogate data can be generated by shuffling the output values (i.e., we permute the outputs keeping fixed the inputs) while keeping fixed the input features. This way we have new data points sharing the same input and output values with the true data, but breaking any structure which links the inputs with the outputs. Clearly, this procedure considers different values for each pair of indices $i$ and $j$ (i.e., each variables $X_i$ and $X_j$). 

Given a set of surrogate data, we applied SCAM-within-MRG with $T=200$ and $M=10$ and obtain the empirical distribution of the hyper-parameter $\delta_{i,j}$. We repeated this procedure $500$ times, generating different surrogate data at each run. We computed different empirical moments of the distribution of the hyper-parameter $\delta_{i,j}$, as mean, median and variance from the empirical distributions obtained via Monte Carlo with the true data and the surrogate data. Averaged results over $500$ runs are shown in Table~\ref{tab1_Ex3_Results}. %, that provides the results corresponding to the mean, median and the standard deviation of the distribution of $\delta_{i,j}$ (averaged over 500 runs). 
We show mean ${\mathbb E}[\delta]$, median ${\bar \delta}_{0.5}$ and standard deviation $\sqrt{\mbox{var}[\delta]}$, obtained analyzing the true data, and the $p$-values obtained comparing the estimated statistics with the corresponding distributions acquired by the surrogate data method. 

\begin{figure}[h!]
\begin{center}
\centerline{
 \subfigure[Mean.]{\includegraphics[width=0.25\textwidth]{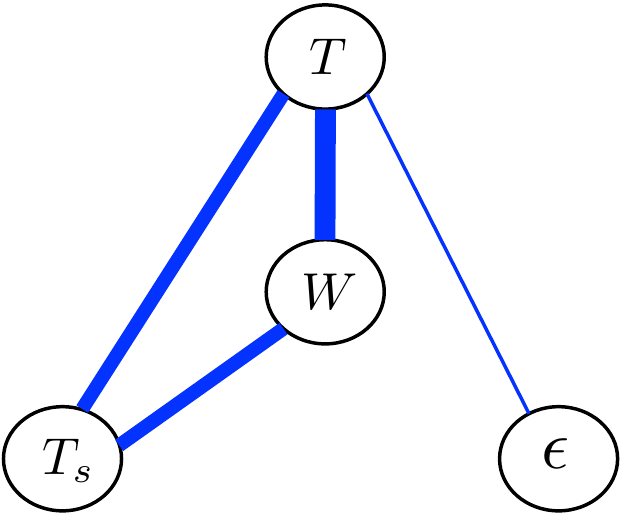}}
 \hspace{0.3cm}
 \subfigure[Median.]{\includegraphics[width=0.25\textwidth]{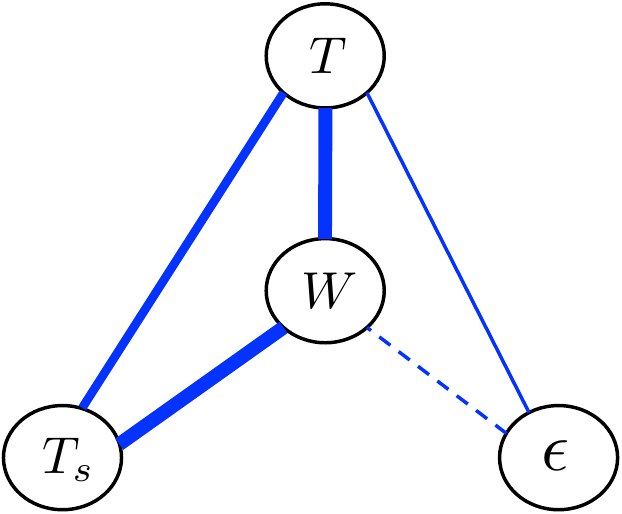}}
  \hspace{0.3cm}
 \subfigure[std.]{\includegraphics[width=0.25\textwidth]{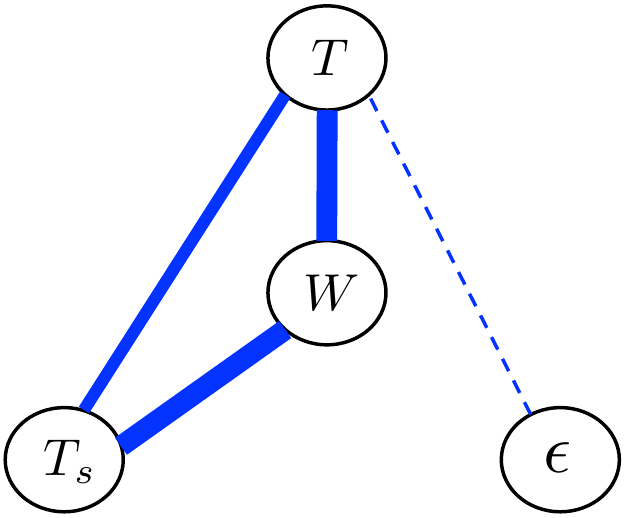}}
}
\caption{{\bf Exp. 3-} Undirected graphs with significant level $\alpha=0.1$. The width of the lines represents the significance of the link, shown in Table~\ref{tab1_Ex3_Results}. If one of the two $p$-values corresponding the two possible regressions is greater than $0.1$ the link is depicted in dashed line.  The graphs are obtained considering {\bf (a)} the mean, {\bf (b)} the median and {\bf (c)} the standard deviation of the distribution of $\delta$. 
}
\label{FigSIMU5}
\end{center}
\end{figure}

Figure~\ref{FigSIMU5} shows the undirected graphs with significance level set to $\alpha=0.1$. The width of the lines represents  the significance of the link according to the estimated $p$-values. If one of the two $p$-values corresponding the two possible regressions between the variable $X_i$ and $X_j$ is greater than $0.1$ the link is depicted in dashed line. The obtained graphs are consistent with a priori physical knowledge about the problem. In particular, it is common sense that the surface temperature $T_s$ and $\epsilon$ depend on $W$ and $T$. After all, remote sensing data processing mainly deals with the estimation of the surface parameters from the acquired satellite brightness temperatures ($T$) and the atmosphere status ($W$)\footnote{While one could be tempted to infer causal relations, the proposed approach cannot cope with asymmetries in the pdfs of variables through GP hyperparameter estimation.}. 
In addition, while emissivity $\epsilon$ is generally used for retrieving $T_s$, some simpler methods using only $T$ are successfully used~\cite{Jimenez07}. It is also worth noting that the used data considered only natural surfaces, hence the database was biased towards high values of $\epsilon$, thus explaining why the relationship between $T_s$ and $\epsilon$ was not captured. %  is nearly independent of $T$ for most common materials and temperature of the terrestrial environment.

 \begin{table}[h!]
%	\centering
%\small
\caption{{\bf Exp. 3-} Results for the mean ${\mathbb E}[ \delta]$, the median ${\bar \delta}_{0.5}$ and the std $\sqrt{\mbox{var}[\delta]}$. }
\centering
\vspace{0.1cm}
\footnotesize
	\begin{tabular}{|c|c|c||c|c||c|c||c|c|}
    \hline
%\begin{enumerate}
%\item
{\bf Link} &{\bf In} & {\bf Out} & {\bf Mean} (${\mathbb E}[\delta]$)  & {\bf$p$-value} & {\bf Median} (${\bm {\bar \delta}_{0.5}}$) & {\bf$p$-value} & {\bf Std} ($\sqrt{\mbox{var}[\delta]}$)  &  {\bf $p$-value} \\
\hline 
\hline 
\multirow{2}{*}{$T_s- W$} & $T_s$& $W$ &  0.68     & 0.004 & 0.67  & 0.002  & 0.15 & 0.001 \\
& $W$& $T_s$ &  0.32  & 0.001  & 0.32  & 0.002  & 0.15 & 0.001 \\
\hline 
\multirow{2}{*}{$T_s- \epsilon$}& $T_s$& $\epsilon$ & 11.85  &  0.22 & 5.61  & 0.20  &  33.16  & 0.56 \\
& $\epsilon$& $T_s$ & 11.58   & 0.23  & 5.67  & 0.23  & 47.36  &   0.68 \\
\hline 
\multirow{2}{*}{$T_s- T$} & $T_s$& $T$ & 2.56  & 0.03  &  2.53   & 0.08  &  0.66 & 0.006\\
& $T$& $T_s$ & 1.83  & 0.02  & 1.76  & 0.03 &  0.48  &0.002 \\
\hline 
\multirow{2}{*}{$W- T$}  & $W$& $T$ &  0.33      &  0.001 &0.34   &  0.001  &  0.14  &0.001\\
& $T$& $W$ &  0.40   & 0.001  & 0.39  & 0.004 &  0.14   & 0.001\\
\hline 
\multirow{2}{*}{$W- \epsilon$} & $W$& $\epsilon$ & 11.77   & 0.22 & 4.04  & 0.09  &32.76  & 0.54\\
& $\epsilon$& $W$ &  11.20 & 0.21  & 4.48  & 0.13  &  47.97  & 0.71 \\
\hline 
\multirow{2}{*}{$T- \epsilon$} & $T$& $\epsilon$ &   5.22 & 0.06  & 3.34  &0.09 &  6.88  & 0.10\\
& $\epsilon$& $T$ & 6.32 &  0.09 & 3.95  & 0.10 &  10.24  & 0.14 \\
\hline 
\end{tabular}
\label{tab1_Ex3_Results}
\end{table}

%%%    

%%%%%%%%%%%%%%%%%%%
\section{Conclusions}
%%%%%%%%%%%%%%%%%%%
\label{ConclSect}

The Gibbs sampling method is a well-known Markov chain Monte Carlo (MCMC) algorithm, extensively applied in statistics, signal processing and machine learning, in order to obtain samples from complicated a posteriori distributions. A Gibbs sampling approach is  particularly useful in high-dimensional inference problems, since the generated samples are constructed component by component. In this sense, it can be considered the MCMC counterpart of the particle filtering methods, for static (i.e., non-sequential inference) and batch (i.e., all the data are processed jointly) frameworks. The key point for the successful application of the SG sampler is the ability to draw efficiently from each the full-conditional densities. However, in real-world applications, drawing from complicated full-conditionals is required, and no direct methods are available in these cases. For solving this issue, several specifically-designed MCMC algorithms has been proposed to be employed within the SG sampler. Most of them require the generation of auxiliary samples that are not included in the resulting estimators. The use of more auxiliary samples accelerates  the convergence of the generated Gibbs chain and improves the performance, at the expense of an additional computational effort. 

In this work, we have shown that these auxiliary samples can be included within the Gibbs estimators improving their efficiency without any extra computational cost. The consistency of the resulting estimators is ensured since the novel MRG scheme is equivalent to an alternative formulation of the well-known chain rule method. This alternative chain rule procedure has been also described and discussed in this work. Numerical simulations have confirmed the benefits of the novel scheme. First, we have compared the SG and MRG schemes in a toy example, considering the use of several parameter values and the application of different internal MCMC algorithms. MRG yielded clear improvements of the performance w.r.t. the SG approach. Then, we tested the SG and MRG schemes in a hyperparameter estimation problem for GP regression, considering a kernel for automatic relevance determination (ARD). % This kernel is often applied in order to analyze the impact of the input components to the outputs of the regression problem.
%We  studied the performance considering many combinations of the parameter values. 
The MRG approach provided the smallest MSEs in estimation of the hyperparameters in all cases. Finally, we studied the application of the proposed MRG sampler to unveil the dependence between different geophysical variables considering remote sensing satellite data. 

As future lines, we plan to investigate the use of different number of samples $M_1$,$...$, $M_d$ to be drawn from full-conditionals, and the possible design of an automatic tuning strategy for adapting the number of samples to improve the performance given a specific posterior distribution. This will imply efforts in parallel MRG samplers for scalable learning. We also plan to extend the numerical study with remote sensing data using the MRG scheme in order to infer causal dependences among the geophysical variables, and for that we will consider applying the MRG in (conditional) independence estimation schemes.

%%%%%%%%%%%%%%%%%%%%%%%%
%\subsection{PMC-like-Gibbs- the monica is related}
%%%%%%%%%%%%%%%%%%%%%%%%%%

\section*{Acknowledgements}
We thank Dr. J. C. Jim\'enez at IPL for the remote sensing dataset and the fruitful discussions. This work has been supported by the European Research Council (ERC) through the ERC Consolidator Grant SEDAL ERC-2014-CoG 647423.

%%%%%%%%%%%%%%%%%
%%%%%%%%%%%%%%%%%
%\bibliographystyle{plainnat}
%\bibliographystyle{plain}
%\bibliographystyle{apalike}
%\bibliographystyle{ims}
%\bibliographystyle{natbib}
%\bibliographystyle{elsarticle-num}
\bibliographystyle{plain}
%\bibliography{pmdrefs,final,pmcrefs,bibliografia}
\bibliography{References}

\begin{thebibliography}{10}

\bibitem{Andrieu2003}
C.~Andrieu, N.~de~Freitas, A.~Doucet, and M.~I. Jordan.
\newblock An introduction to {MCMC} for machine learning.
\newblock {\em Machine Learning}, 50(1):5--43, 2003.

\bibitem{Bishop}
C.~Bishop.
\newblock {\em {P}attern {R}ecognition and {M}achine {L}earning}.
\newblock Springer, 2006.

\bibitem{Brewer93}
M.~Brewer and C.~Aitken.
\newblock Discussion on the meeting on the {G}ibbs sampler and other {M}arkov
  {C}hain {M}onte {C}arlo methods.
\newblock {\em Journal of the Royal Statistical Society. Series B},
  55(1):69--70, 1993.

\bibitem{Bugallo07}
M.~F. Bugallo, S.~Xu, and P.~M. Djuri\'c.
\newblock Performance comparison of {EKF} and particle filtering methods for
  maneuvering targets.
\newblock {\em Digital Signal Processing}, 17:774--786, October 2007.

\bibitem{Caffo02}
B.~S. Caffo, J.~G. Booth, and A.~C. Davison.
\newblock Empirical supremum rejection sampling.
\newblock {\em Biometrika}, 89(4):745--754, December 2002.

\bibitem{Cai08}
B.~Cai, R.~Meyer, and F.~Perron.
\newblock Metropolis-{H}astings algorithms with adaptive proposals.
\newblock {\em Statistics and Computing}, 18:421--433, 2008.

\bibitem{Chen16}
Y.~Chen, L.~Bornn, N.~De~Freitas, M.~Eskelin, J.~Fang, and M.~Welling.
\newblock Herded {G}ibbs sampling.
\newblock {\em Journal of Machine Learning Research}, 17(1):263--291, 2016.

\bibitem{Devroye86}
L.~Devroye.
\newblock {\em Non-Uniform Random Variate Generation}.
\newblock Springer, 1986.

\bibitem{Djuric03}
P.~M. Djuri\'c, J.~H. Kotecha, J.~Zhang, Y.~Huang, T.~Ghirmai, M.~F. Bugallo,
  and J.~M\'{\i}guez.
\newblock Particle filtering.
\newblock {\em IEEE Signal Processing Magazine}, 20(5):19--38, September 2003.

\bibitem{Fitzgerald01}
W.~J. Fitzgerald.
\newblock {M}arkov chain {M}onte {C}arlo methods with applications to signal
  processing.
\newblock {\em Signal Processing}, 81(1):3--18, January 2001.

\bibitem{Fox12}
C.~Fox.
\newblock A {G}ibbs sampler for conductivity imaging and other inverse
  problems.
\newblock {\em Proc. of SPIE, Image Reconstruction from Incomplete Data VII},
  8500:1--6, 2012.

\bibitem{Gelfand93}
A.~E. Gelfand and T.~M. Lee.
\newblock Discussion on the meeting on the {G}ibbs sampler and other {M}arkov
  {C}hain {M}onte {C}arlo methods.
\newblock {\em Journal of the Royal Statistical Society. Series B},
  55(1):72--73, 1993.

\bibitem{Gilks92derfree}
W.~R. Gilks.
\newblock {D}erivative-free adaptive rejection sampling for {G}ibbs sampling.
\newblock {\em {B}ayesian Statistics}, 4:641--649, 1992.

\bibitem{Gilks95}
W.~R. Gilks, N.~G. Best, and K.~K.~C. Tan.
\newblock {A}daptive rejection {M}etropolis sampling within {G}ibbs sampling.
\newblock {\em Applied Statistics}, 44(4):455--472, 1995.

\bibitem{CorrGilks97}
W.~R. Gilks, R.M. Neal, N.~G. Best, and K.~K.~C. Tan.
\newblock Corrigidum: {A}daptive rejection {M}etropolis sampling within {G}ibbs
  sampling.
\newblock {\em Applied Statistics}, 46(4):541--542, 1997.

\bibitem{Gilks92}
W.~R. Gilks and P.~Wild.
\newblock {A}daptive rejection sampling for {G}ibbs sampling.
\newblock {\em Applied Statistics}, 41(2):337--348, 1992.

\bibitem{Gorur08rev}
Dilan G{\"o}r{\"u}r and Yee~Whye Teh.
\newblock Concave convex adaptive rejection sampling.
\newblock {\em Journal of Computational and Graphical Statistics},
  20(3):670--691, September 2011.

\bibitem{Goudie16}
R.~J.~B. Goudie and S.~Mukherjee.
\newblock A {G}ibbs sampler for learning {DAG}s.
\newblock {\em Journal of Machine Learning Research}, 17(2):1--39, 2016.

\bibitem{Haario01}
H.~Haario, E.~Saksman, and J.~Tamminen.
\newblock An adaptive {M}etropolis algorithm.
\newblock {\em Bernoulli}, 7(2):223--242, April 2001.

\bibitem{HaarioCW}
H.~Haario, E.~Saksman, and J.~Tamminen.
\newblock Component-wise adaptation for high dimensional {MCMC}.
\newblock {\em Computational Statistics}, 20(2):265--273, 2005.

\bibitem{Hoermann95}
W.~H{\"o}rmann.
\newblock A rejection technique for sampling from {T}-concave distributions.
\newblock {\em ACM Transactions on Mathematical Software}, 21(2):182--193,
  1995.

\bibitem{Hormann02}
W.~H{\"o}rmann.
\newblock A note on the performance of the {A}hrens algorithm.
\newblock {\em Computing}, 69:83--89, 2002.

\bibitem{Hormann07}
W.~H{\"o}rmann, J.~Leydold, and G.~Derflinger.
\newblock Inverse transformed density rejection for unbounded monotone
  densities.
\newblock {\em Research Report Series/ Department of Statistics and Mathematics
  (Economy and Business), Vienna University}, 2007.

\bibitem{Jimenez07}
J.~C. Jimenez-Munoz and J.~A. Sobrino.
\newblock Feasibility of retrieving land-surface temperature from aster tir
  bands using two-channel algorithms: A case study of agricultural areas.
\newblock {\em IEEE Geoscience and Remote Sensing Letters}, 4(1):60--64, Jan
  2007.

\bibitem{Jimenez08}
J.~C. Jimenez-Mu{\~n}oz and J.~A. Sobrino.
\newblock Split-window coefficients for land surface temperature retrieval from
  low resolution thermal infrared sensors.
\newblock {\em IEEE Geoscience and Remote Sensing Letters}, 5(4):806--809,
  2008.

\bibitem{Johnson13}
A.~A. Johnson, G.~L. Jones, and R.~C. Neath.
\newblock Component-wise {M}arkov {C}hain {M}onte {C}arlo: uniform and
  geometric ergodicity under mixing and composition.
\newblock {\em Statistical Science}, 28(3):360--375, 2013.

\bibitem{Koch07}
K.~R. Koch.
\newblock Gibbs sampler by sampling-importance-resampling.
\newblock {\em Journal of Geodesy}, 81(9):581--591, 2007.

\bibitem{Kotecha99}
J.~Kotecha and Petar~M. Djuri\'c.
\newblock Gibbs sampling approach for generation of truncated multivariate
  {G}aussian random variables.
\newblock {\em Proceedings of Acoustics, Speech, and Signal Processing,
  (ICASSP)}, 1999.

\bibitem{Levine05}
R.~A. Levine, Z.~Yu, W.~G. Hanley, and J.~J. Nitao.
\newblock Implementing component-wise {H}astings algorithms.
\newblock {\em Computational Statistics and Data Analysis}, 48(2):363--389,
  2005.

\bibitem{Liang10}
F.~Liang, C.~Liu, and R.~Caroll.
\newblock {\em Advanced {M}arkov {C}hain {M}onte {C}arlo Methods: Learning from
  Past Samples}.
\newblock Wiley Series in Computational Statistics, England, 2010.

\bibitem{Liu04b}
J.~S. Liu.
\newblock {\em {M}onte {C}arlo Strategies in Scientific Computing}.
\newblock Springer-Verlag, 2004.

\bibitem{Lucka16}
F.~Lucka.
\newblock Fast {G}ibbs sampling for high-dimensional {B}ayesian inversion.
\newblock {\em arXiv:1602.08595}, 2016.

\bibitem{Marrelec04}
G.~Marrelec and H.~Benali.
\newblock Automated rejection sampling from product of distributions.
\newblock {\em Computational Statistics}, 19(2):301--315, May 2004.

\bibitem{PARS}
L.~Martino.
\newblock Parsimonious adaptive rejection sampling.
\newblock {\em IET Electronics Letters}, 53(16):1115--1117, 2017.

\bibitem{Sticky13}
L.~Martino, R.~Casarin, F.~Leisen, and D.~Luengo.
\newblock Adaptive independent sticky {MCMC} algorithms.
\newblock {\em (to appear) EURASIP Journal on Advances in Signal Processing},
  pages 1--32, 2017.

\bibitem{CARS}
L.~Martino and F.~Louzada.
\newblock Adaptive rejection sampling with fixed number of nodes.
\newblock {\em (To appear) Communications in Statistics - Simulation and
  Computation (DOI: 10.1080/03610918.2017.1395039)}, pages 1--11, 2017.

\bibitem{MartinoStatCo10}
L.~Martino and J.~M\'{\i}guez.
\newblock A generalization of the adaptive rejection sampling algorithm.
\newblock {\em Statistics and Computing}, 21(4):633--647, October 2011.

\bibitem{MartinoA2RMS}
L.~Martino, J.~Read, and D.~Luengo.
\newblock Independent doubly adaptive rejection {M}etropolis sampling within
  {G}ibbs sampling.
\newblock {\em IEEE Transactions on Signal Processing}, 63(12):3123--3138, June
  2015.

\bibitem{FUSS}
L.~Martino, H.~Yang, D.~Luengo, J.~Kanniainen, and J.~Corander.
\newblock A fast universal self-tuned sampler within {G}ibbs sampling.
\newblock {\em Digital Signal Processing}, 47:68 -- 83, 2015.

\bibitem{Meyer08}
R.~Meyer, B.~Cai, and F.~Perron.
\newblock Adaptive rejection {M}etropolis sampling using {L}agrange
  interpolation polynomials of degree 2.
\newblock {\em Computational Statistics and Data Analysis}, 52(7):3408--3423,
  March 2008.

\bibitem{Muller91}
P.~M{\"u}ller.
\newblock A generic approach to posterior integration and {G}ibbs sampling.
\newblock {\em Technical Report 91-09, Department of Statistics of Purdue
  University}, 1991.

\bibitem{rasmussen2006gaussian}
C.~E. Rasmussen and C.~K.~I. Williams.
\newblock {\em Gaussian processes for machine learning}.
\newblock MIT Press, 2006.

\bibitem{ReadLuca2014}
J.~Read, L.~Martino, and D.~Luengo.
\newblock Efficient {M}onte {C}arlo methods for multi-dimensional learning with
  classifier chains.
\newblock {\em Pattern Recognition}, 47(3):1535 -- 1546, 2014.

\bibitem{ritter1992griddyGibbs}
C.~Ritter and M.~A. Tanner.
\newblock Facilitating the {G}ibbs sampler: The {G}ibbs stopper and the
  griddy-{G}ibbs sampler.
\newblock {\em Journal of the American Statistical Association},
  87(419):861--868, 1992.

\bibitem{Robert04}
C.~P. Robert and G.~Casella.
\newblock {\em {M}onte {C}arlo Statistical Methods}.
\newblock Springer, 2004.

\bibitem{roberts1997updating}
G.~O. Roberts and S.~K. Sahu.
\newblock Updating schemes, correlation structure, blocking and
  parameterization for the {G}ibbs sampler.
\newblock {\em Journal of the Royal Statistical Society: Series B},
  59(2):291--317, 1997.

\bibitem{Shao13}
W.~Shao, G.~Guo, F.~Meng, and S.~Jia.
\newblock An efficient proposal distribution for {M}etropolis-{H}astings using
  a b-splines technique.
\newblock {\em Computational Statistics and Data Analysis}, 53:465--478, 2013.

\bibitem{Sobrino08}
J.~A. Sobrino, J.~C. Jimenez-Mu{\~n}oz, G.~Soria, M.~Romaguera, L.~Guanter,
  J.~Moreno, A.~Plaza, and P.~Martinez.
\newblock Land surface emissivity retrieval from different {VNIR} and {TIR}
  sensors.
\newblock {\em IEEE Transactions on Geoscience and Remote Sensing},
  46(2):316--327, 2008.

\bibitem{Tanizaki99}
H.~Tanizaki.
\newblock On the nonlinear and non-normal filter using rejection sampling.
\newblock {\em {IEEE} {T}ransaction on automatic control}, 44(3):314--319,
  February 1999.

\bibitem{Theiler92}
J.~Theiler, S.~Eubank, A.~Longtin, B.~Galdrikian, and J.~D. Farmer.
\newblock Testing for nonlinearity in time series: the method of surrogate
  data.
\newblock {\em Physica D}, 58(2):77--94, 1992.

\bibitem{Zhang16}
H.~Zhang, Y.~Wu, L.~Cheng, and I.~Kim.
\newblock Hit and run {ARMS}: adaptive rejection {M}etropolis sampling with hit
  and run random direction.
\newblock {\em Journal of Statistical Computation and Simulation},
  86(5):973--985, 2016.

\end{thebibliography}

%This work has been supported by the Grant 2014/23160-6 of S\~ao Paulo Research Foundation (FAPESP). %and by the Grant 305361/2013-3 of National Council for Scientific and Technological Development (CNPq).

%\bibliographystyle{elsarticle-num} 
%\bibliographystyle{IEEEtran}
%\bibliography{bibliografia,biblioFading}
%\bibliography{bib_battery,other,traffic_new,bibliografia}

%\bibliography{bibliografia}

\end{document}